\documentclass[preprint,aps,prd,showpacs,nofootinbib,superscriptaddress,floatfix,showkeys]{revtex4-2}

\usepackage{amsfonts}
\usepackage{amsmath}
\usepackage{amssymb} 
\usepackage{bm} 
\usepackage{color}
\usepackage{enumerate}
\usepackage{graphicx}  
\usepackage{dcolumn}
\usepackage{hyperref}
\usepackage{lineno}

\hypersetup{
	colorlinks=true,       
	linkcolor=blue,          
	citecolor=magenta,        
	filecolor=cyan,      
	urlcolor=magenta           
}

\labelformat{section}{#1} 
\labelformat{subsection}{Section #1} 
\labelformat{figure}{#1} 
\labelformat{subfigure}{Fig.~\thefigure#1} 
\labelformat{table}{Tab.~#1} 

\def\be{\begin{equation}}
	\def\ee{\end{equation}}
\def\ben{\begin{eqnarray}}
	\def\een{\end{eqnarray}}

\begin{document}
\title{Revisiting Constraints on Primordial Curvature Power Spectrum from PBH Abundances} 
\author{Ashu Kushwaha} 
\author{Teruaki Suyama}
\affiliation{Department of Physics, Institute of Science Tokyo, 2-12-1 Ookayama, Meguro-ku, Tokyo 152-8551, Japan}
%
	%
	%
\begin{abstract}
Primordial black holes (PBHs) can form in the early Universe, for instance during radiation domination, from the collapse of large-amplitude density perturbations shortly after horizon re-entry. This mechanism establishes an approximate one-to-one correspondence between the PBH mass and the scale of the peak in the primordial curvature perturbations. Consequently, the constraints on PBH abundances can be translated into upper limits on the amplitude of the primordial curvature power spectrum, thereby providing an indirect probe of the last e-folds of inflation corresponding to these smaller scales. Assuming Gaussian primordial curvature perturbations, we derive 
constraints on the amplitude of primordial curvature power spectra with both narrow and broad peaks using the most up-to-date bounds on PBH abundances. Given the theoretical uncertainties in PBH formation, we systematically compare the constraints obtained using the Press–Schechter (PS) formalism and peak theory, accounting for the nonlinear relation between curvature perturbations and density contrast, and assess the impact of different window functions and the transfer function. We quantify the impact of spherical versus non-spherical collapse criteria and show that including non-sphericity significantly increases the inferred amplitude of the primordial power spectrum, reflecting the larger threshold density contrast required for PBH formation. We also find that whereas the constraints obtained using the PS formalism and peak theory remain largely similar for the monochromatic case, they differ in the case of a broad primordial power spectrum. 
This discrepancy underscores that current constraints remain sensitive to the choice of statistical formalism. 
Our consistent treatment of monochromatic and extended mass functions provides a systematic 
mapping based on existing methodologies, 
while highlighting that reducing these theoretical uncertainties is a crucial step toward 
probing the early Universe through PBHs.  

\end{abstract}
\maketitle
\newpage

\section{Introduction}

Cosmic microwave background (CMB) and large-scale structure (LSS) observations have established the amplitude of primordial curvature perturbations on cosmological scales with high precision, 
thereby placing strong constraints on inflationary models~\cite{Book-Dodelson-2020,Book-Baumann-2022}. 
These measurements, however, probe only a limited range of comoving scales (for example, $10^{-4} \, {\rm Mpc}^{-1} \lesssim k \lesssim 0.01 \, {\rm Mpc}^{-1} $), 
leaving the properties of primordial perturbations on much smaller scales 
essentially unconstrained by direct observations.

Primordial fluctuations on small scales are nevertheless of particular interest, since they encode information about the final stages of inflation. 
At present, the most stringent constraints on such small-scale perturbations arise from the non-detection of primordial black holes (PBHs)\footnote{Other stringent small-scale constraints come from CMB distortions~\cite{Byrnes:2018txb,Inomata:2018epa} and ultracompact minihalos~\cite{Bringmann:2025cht}.}. 
If the amplitude of primordial fluctuations were sufficiently large (which can be achieved by introducing nontrivial physics during inflation, see Refs.~\cite{Garcia-Bellido:1996mdl,Yokoyama:1998pt,Kohri:2007qn,Kawasaki:2012wr,Garcia-Bellido:2017mdw,Inomata:2017okj,Carr:2018nkm,Byrnes:2018txb,Inomata:2018cht,Bhaumik:2019tvl,Braglia:2020eai,Ragavendra:2020sop}), 
PBHs would be produced abundantly in the early Universe~\cite{Zeldovich:1967lct,1971-Hawking-MNRAS,Carr:1974nx,Chapline:1975ojl,Khlopov:2008qy,2016-Carr.Kuhnel.Sandstad-PRD,2018-Sasaki.etal-CQG,2021-Carr.etal-Rept.Prog.Phys,2020-Green.Kavanagh-JPhyG,Escriva:2022duf,Domenech:2026nun}. 
The absence of observational evidence for PBHs therefore allows one to place upper bounds on the amplitude of the primordial curvature power spectrum on scales 
far smaller than those accessible to CMB and LSS observations.

Since PBHs are formed shortly after the horizon re-entry during the RD epoch, this provides an approximate one-to-one correspondence between the mass of the PBHs and the scale of the peak in the primordial curvature perturbations during inflation. Interestingly, this connection allows us to translate the constraints from one to the other; for instance, we can constrain the primordial curvature power spectrum from the PBHs abundance, which is the primary focus of this work\footnote{Similarly, we can constrain the power spectrum of primordial magnetic fields during the RD epoch from the abundance of PBHs, see Ref.\cite{Kushwaha:2024zhd} for more details.}. This method has been investigated in the literature~\cite{Sato-Polito:2019hws,Kalaja:2019uju,Gow:2020bzo}. Due to the uncertainties associated with the PBHs' formation mechanisms, the constraints on the primordial power spectrum are also affected, which requires more refined calculations by a systematic treatment of potential sources of uncertainties. For example, the formation of PBHs has been broadly discussed within the context of Press-Schechter (PS) formalism~\cite{1974-Press-Schechter-APJ} and Peak theory~\cite{Bardeen:1985tr}. However, the estimation of the PBHs mass distribution (or the abundance of PBHs) differs between the two approaches~\cite{Young:2014ana,2018-Sasaki.etal-CQG,Yoo:2020dkz,Young:2020xmk}, and the preference for one over the other remains uncertain. Another important source of uncertainty is the correct prescription of the shape of the overdense region, which is often considered to be spherical; however, it is shown that the threshold density contrast increases for an ellipsoidal overdense region, thus affecting the constraints~\cite{Kuhnel:2016exn}.

In this work, we revisit the study of constraining the small-scale primordial curvature power spectrum from the PBHs abundance. By using the most updated constraints on PBHs abundance provided in Ref.~\cite{2021-Carr.etal-Rept.Prog.Phys}, we perform a systematic study of various issues leading to the uncertainties in the estimation of PBHs abundance. Specifically, using the refined calculations based on the compaction function, with the non-linear relation between density contrast and curvature perturbation and including non-sphericity of the overdense region, we compare the constraints for the following cases: (i) PS formalism with Peak theory, (ii) monochromatic and extended mass distribution of PBHs, and (iii) different choices of window functions, including the role of the transfer function in the computation of the density variance. Throughout this work, we assume that the primordial curvature perturbations obey Gaussian statistics. Since PBHs form from rare fluctuations in the tail of the distribution, even a small local-type non-Gaussianity can change the PBH abundance by orders of magnitude~\cite{Byrnes:2012yx,Franciolini:2018vbk,Atal:2019cdz,Inui:2024fgk}; the constraints derived here should therefore be understood as applying to the Gaussian case. We highlight the impact of these choices on the resulting constraints and discuss the associated theoretical uncertainties. Note that some of these effects have been considered in the previous literature (see Refs.~\cite{Zaballa:2006kh,Bugaev:2008gw,Sato-Polito:2019hws,Kalaja:2019uju,Gow:2020bzo,Yi:2022ymw}, however, to the best of our knowledge, all these effects have not been considered together to derive the constraints on the primordial curvature power spectrum. Therefore, the constraints obtained in this work may provide a better comparison and can be considered as more precise and robust.
This work is organized in the following way: in section~\ref{sec-comp-func}, we discuss the criterion for the collapse based on the compaction function. Section~\ref{sec-pbh-computation} discusses the methodology to compute the PBHs abundance, and in section~\ref{sec-constraint} we provide the final constraints on the power spectrum. Unless otherwise stated, the results shown in the main text are obtained using a real-space top-hat window function together with the transfer function. The impact of different choices of window function and the inclusion of the transfer function are discussed in Appendix~\ref{appsec-window-fun}, while Appendix~\ref{appsec:constraints-tophat} presents the corresponding constraints obtained using the Gaussian window function. Section~\ref{sec-conclusion} provides the summary.

\section{The compaction function and criterion for the collapse of an overdense region: A brief review}\label{sec-comp-func}
In this section, we provide a brief review of the criterion for the collapse of an overdense region into PBHs, based on earlier studies~\cite{2019-Musco-PRD,Young:2019yug,Kalaja:2019uju}, which will be used in the subsequent analysis. Let us consider a spacetime where superhorizon-scale spherically symmetric overdense region is superposed on top of the flat Friedmann–Lemaître–Robertson–Walker (FLRW) universe.
The metric of this spacetime is given by
\begin{align}\label{metric-curvature}
	ds^2 = -dt^2 + a^2(t) e^{2\zeta (r)} \left[ dr^2 + r^2 d\Omega^2\right]
\end{align}
where $a(t)$ is the scale factor in cosmic time $t$, $\zeta(r)$ is the curvature perturbation, $r$ is the radial coordinate which is related to 
the physical radius (or areal radius) as $R(r,t) = a(t) r e^{\zeta (r) } $.
The local overdensity (or density contrast) of a region of radius $r$ centered at a point $\textbf{x}$ is defined as $\frac{\delta \rho (r,t,\textbf{x})}{\bar{\rho}(t)} = \frac{\rho (r,t,\textbf{x}) - \bar{\rho}(t)}{\bar{\rho}(t)} $, where $\rho (r,t,\textbf{x}) $ is the energy density of the overdense region, $\bar{\rho} (t) = \frac{3H^2}{8\pi}$ is the background energy density and $H(t) = \dot{a}/a$ is the Hubble parameter.

In the gradient expansion formalism, the non-linear relation between the density contrast $\delta \rho/\bar{\rho}$ (defined on a comoving uniform-cosmic time slicing) and the curvature perturbation $\zeta$ for a spherically symmetric configuration is given by~\cite{Harada:2015yda,2019-Germani.Musco-PRL,2019-Musco-PRD,Young:2019yug,Kalaja:2019uju}
\begin{align}\label{delta-rho-nl-flrw}
	\frac{\delta \rho}{\bar{\rho}}  = -  \frac{2 (1+w)}{5+3w} \left(\frac{1}{aH}\right)^2 e^{-2\zeta (r)}  \left[ \zeta'' (r) + \frac{2}{r} \zeta'(r) + \frac{1}{2} {\zeta'(r)}^2 \right]~,
\end{align}
where $w = p/\rho$ is the equation of the state parameter, which is $1/3$ for the RD era, and ${}^{\prime} \equiv d/dr$. At linear order, the above equation gives
\begin{align}\label{delta-rho-l-flrw}
	\frac{\delta \rho}{\bar{\rho}}  \simeq -  \frac{2 (1+w)}{5+3w} \left(\frac{1}{aH}\right)^2 \left[ \zeta'' (r) + \frac{2}{r} \zeta'(r) \right] ~~.
\end{align}
Note that the density contrast $\delta \rho/\bar{\rho}$ is a local quantity and is defined at a given location.
However, PBH formation is determined by whether the region as a whole is sufficiently overdense when it re-enters the horizon, which depends on the total mass excess enclosed within a finite volume rather than on the local density contrast. It is therefore more useful to consider the volume-averaged density contrast within a sphere of areal radius $R$, which directly captures this condition. This quantity is defined as
\begin{align}\label{delta-def}
	\delta (r,t) = \frac{1}{V} \int_0^R dR \, 4\pi R^2 \frac{\delta \rho}{\bar{\rho}}
\end{align}
where $V = 4\pi R^3/3$. 

\subsection{Equivalence between compaction function and averaged (smoothed) density contrast}

In recent years, the compaction function has become a popular quantity in calculating PBH abundance. As compared to the average density contrast, the compaction function allows
us to consistently determine the correct radius (length scale) of the overdense region and compare the curvature and overdensity profiles with different shapes. The compaction function quantifies the magnitude of the gravitational potential and is defined as twice the ratio between the mass excess $\delta M$ inside a sphere of areal radius $R(r,t)$ at a time $t$ to the areal radius itself  as\footnote{The compaction function was introduced by Shibata and Sasaki~\cite{Shibata:1999zs} as $C = \delta M/R$.}~\cite{Musco:2018rwt,Kalaja:2019uju,Young:2019yug,Harada:2023ffo,Harada:2024trx,Young:2024jsu} 
\begin{align}\label{compaction-fun-def}
	C (r,t,\textbf{x}) \equiv \frac{2 \delta M (r,t,\textbf{x}) }{R(r,t,\textbf{x})}~~,
\end{align}
where $\delta M (r,t,\textbf{x}) = M (r,t,\textbf{x}) - \bar{M} (t)$ is the excess mass within a sphere of areal radius $ R(t,r,\textbf{x})$ centered at the point $\textbf{x}$ and $\bar{M}(t) = \frac{4\pi R^3(r,t)}{3}\bar{\rho}(t)$ is the mass corresponding to the background energy density. One important feature of the compaction function is that it is conserved on superhorizon scales, where $H(t) R(r,t) >>1$. Therefore, we can take the compaction function as a time-independent quantity before the horizon crossing time $t_{hc}$. 
The length scale of the perturbation (i.e., the radius of the overdense region) can be determined by the location $r_m$ of the local maxima of the compaction function as $C'(r_m) = 0$, see Refs.~\cite{Musco:2018rwt,Young:2019yug}. 
Because the maxima of the compaction function (measured by $2M/R$) directly corresponds to the maxima of the Newtonian gravitational potential (measured by $M/R$), $r_m$ gives the correct and physically well-motivated estimation of the radius of the overdense region. 

Let us first calculate the mass excess as the volume integral of excess density with a sphere of physical radius $R_m$ corresponding to the location of maxima of the compaction function i.e., $r_m$ as
\begin{align}
	\delta M (R_m,t) = M(R_m,t) - \bar{M} (t) & = \bar{\rho} \int_0^{R_m} dR \,  4\pi R^2 \frac{\delta \rho }{\bar{\rho}} ~.
\end{align}
Using the non-linear relation~\eqref{delta-rho-nl-flrw} for $\delta \rho / \bar{\rho}$ in the above equation gives the compaction function~\eqref{compaction-fun-def} for the non-linear relation as~\cite{Musco:2018rwt,Kalaja:2019uju}
\begin{align}\label{cnl-final}
	C_{\rm NL} (r_m) = - \frac{3 (1+w)}{5+3w} r_m \zeta' (r_m)  \left[ 2 + r_m \zeta'(r_m) \right] ~~.
\end{align}
Similarly, the compaction function for the linear relation~\eqref{delta-rho-l-flrw}, can be given as
\begin{align}\label{cl-final}
	C_L (r_m) = - \frac{6 (1+w)}{5+3w} r_m \zeta' (r_m) ~~.
\end{align}
Note that the subscripts ${}_L$ and ${}_{NL}$ refer to the linear~\eqref{delta-rho-l-flrw} and non-linear relations~\eqref{delta-rho-nl-flrw} between the curvature perutbration $\zeta$ and density contrast $\delta \rho/\bar{\rho}$.
Using Eq.\eqref{cl-final} in Eq.\eqref{cnl-final} allows us to obtain the non-linear compaction function in terms of the linear component as
\begin{align}\label{cnl-cl-relation}
	C_{\rm NL} = C_L - \frac{1}{12}\left(\frac{5+3w}{1+w} \right)  C_L^2 = C_L - \frac{3}{8}  C_L^2
\end{align}
where in the last expression we used $w=1/3$ for the RD era. 
The above equation suggests that $C_{\rm NL}$ is maximum at $C_L = 4/3$ with $C_{\rm NL, max} = 2/3$.

From the definitions of $\delta$ and $C$, it follows that $C_{\rm NL}(r_m)=H^2 R_m^2 \delta (r_m,t)$.
Thus, we can identify $C_{\rm NL}(r_m)$ with the smoothed density contrast evaluated
at the horizon reentry time $HR_m=1$. This identification is, however, only approximate, since it neglects the evolution of perturbations near the horizon crossing. A more accurate treatment would require the inclusion of the transfer function describing this evolution. Nevertheless, one can verify that, for the comoving density contrast $\delta(r_m,t)$, the value obtained at horizon-crossing differs from its leading-order super-horizon\footnote{
	During RD era, the linear transfer function for the comoving density contrast is $T(x) \propto \frac{ \sin(x) - x \cos(x)}{x}$, where $x=k\eta/\sqrt{3}$, whereas the leading-order gradient expansion gives $T(x)\propto x^2/3$ for $x\ll1$ (super-horizon limit). Evaluating both expressions at horizon crossing ($x=1/\sqrt3$) shows that they differ by only $\sim3$--$4\%$, indicating that the leading-order super-horizon approximation remains an excellent estimate of the density contrast at horizon crossing. We note, however, that this estimate may depend on the specific density profile and the choice of smoothing window function. A detailed investigation of these effects is beyond the scope of the present work.} value by only $\sim 3$--$4\%$. Since this correction is small, we adopt the above identification throughout this work for simplicity.
Denoting by $\delta_L$ the linear part of $\delta_m (r_m,t_{\rm hc})$, 
where $t_{\rm hc}$ is the horizon reentry time of the scale $r_m$,
Eq.~(\ref{cnl-cl-relation}) gives
\begin{align}\label{deltam-deltal-relation}
	\delta_m =  \delta_L - \frac{3}{8}  \delta_L^2, ~~
\end{align}
where $\delta_m \equiv \delta_m (r_m,t_{\rm hc})$.
The LHS of the above equation can be associated with the threshold density contrast of PBHs formation ($\delta_c$), and we will refer this as $\delta_{m,c}$. 
Solving for the roots of the above quadratic equation gives the corresponding threshold value of the linear component $\delta_{L,c}$ as
\begin{align}\label{delta-lc-roots}
	\delta_{L\pm,c} = \frac{4}{3} \left( 1 \pm \sqrt{1 - \frac{3}{2}\delta_{m,c}} \right) ~.
\end{align}
One can see that for a typical values of the threshold density contrast, $\delta_c = 0.45 = \delta_{m,c}$ for the real-space top-hat window function (or $0.18$ for the Gaussian window function)~\cite{Young:2019osy}, the corresponding root is $\delta_{L+,c} = \frac{4}{3} \left( 1 + \sqrt{1 - \frac{3}{2}\delta_{m,c}} \right) \simeq 2.09$ (or $2.47$ for the Gaussian case) corresponds to very large amplitude perturbations referred to as the type-II perturbations.
For Gaussian primordial fluctuations, type-II fluctuations are expected to form PBHs; however, since they require extremely large amplitudes, their contribution to the PBH abundance is exponentially suppressed\footnote{The situation can differ for non-Gaussian statistics, where type-II fluctuations may contribute significantly to the abundance, and their fate depends on the form of non-Gaussianity: they collapse into PBHs in the set-up of Ref.~\cite{Uehara:2024yyp}, while for strongly negative logarithmic non-Gaussianity the formation threshold can fall within the type-II regime, so that not all of them form PBHs~\cite{Inui:2024fgk}.}. We therefore restrict our analysis to type-I fluctuations and consider the physically relevant root $\delta_{L-,c}$.
This threshold incorporates the nonlinear relation between the density contrast and the curvature perturbation while allowing us to work with the linear density contrast $\delta_L$. Because $\delta_L$ is linearly related to the curvature perturbation $\zeta$, 
the statistical properties of $\delta_L$ are the same as those of $\zeta$(although this correspondence is not exact for derived quantities such as the variance). For later purpose, we provide the linear relation in Fourier space obtained by combining
Eq.~(\ref{delta-rho-l-flrw}) and the linearized Eq.~(\ref{delta-def}):
\begin{equation}
	\label{relation-deltaL-zeta}
	{\tilde \delta_L}({\bm k},t)=\frac{2(1+w)}{5+3w} \frac{k^2}{a^2H^2}
	{\tilde W}(k,R) {\tilde \zeta}_{\bm k},
\end{equation}
where ${\tilde W}(k,r_m)$ is the Fourier transform of the window function. 
In the main part of this work, we adopt the real-space top-hat window function
	%
	\begin{align}\label{th}
		\tilde{W}_{\rm TH}(k,r_m) = 3\frac{\sin{(kr_m)} - kr_m \cos{(kr_m)}}{(kr_m)^3} = 3\frac{j_1 (kr_m)}{kr_m},
	\end{align}
	and $j_1$ is the spherical Bessel function. 
The choice of window function introduces a systematic uncertainty in the calculation of the PBH abundance and, consequently, in the inferred constraints on the primordial power spectrum. Besides the real-space top-hat window function, another commonly used choice is the Gaussian window function, 
	whose Fourier space is
	\begin{align}
		\tilde{W}_G(k,r_m) = \exp\left( -\frac{(k\,r_m)^2}{4} \right)~.
	\end{align}
From a physical perspective, the real-space top-hat window function provides a natural smoothing prescription for PBH formation in the sense that it completely excludes contributions from $\zeta$ beyond the Hubble horizon.
However, its oscillatory behavior in Fourier space leads to significant numerical challenges in the evaluation of the variance and related peak-theory quantities. These difficulties become particularly severe for broad primordial power spectra, while they are less pronounced for narrow spectra.
To account for the evolution of perturbations after horizon re-entry and to alleviate these numerical issues, one commonly introduces the radiation-era transfer function,
\begin{align}
	T (k,r_m) = 3\frac{\sin{\left(\frac{kr_m}{\sqrt{3}} \right)} - \left(\frac{kr_m}{\sqrt{3}} \right) \cos{\left(\frac{kr_m}{\sqrt{3}} \right)}}{\left(\frac{kr_m}{\sqrt{3}} \right)^3} = 3\frac{j_1 \left(\frac{kr_m}{\sqrt{3}} \right)}{\left(\frac{kr_m}{\sqrt{3}} \right)}~~.
\end{align}
This transfer function accounts for the sub-horizon evolution of radiation perturbations and suppresses contributions from modes well inside the horizon.
From a physical perspective, its inclusion in the computation of the PBH abundance is well motivated. In particular, for broad primordial power spectra, the sub-horizon evolution encoded in the transfer function can have a non-negligible impact on the predicted PBH abundance, regardless of the choice of window function.
As shown in appendix~\ref{appsec-window-fun}, the inclusion of the transfer function modifies the resulting constraints by up to ($\sim 30\%$) when a Gaussian window function is employed, while its impact is considerably larger for the real-space top-hat window function. A detailed comparison of the constraints obtained for broad primordial power spectra, with and without the transfer function, is presented in Appendix~\ref{appsec-window-fun}.
Unless otherwise stated, all results presented in the main text are computed using the real-space top-hat window function~\eqref{th} together with the transfer function defined above.

Before proceeding, we emphasize that the derivation presented above is valid for the real-space top-hat window function (which is used in the main text). However, in Appendix~\ref{appsec:constraints-tophat} we use the same derivation for a Gaussian window function as well, for which both the smoothing of the non-linear density contrast and the resulting probability distribution of the compaction function are modified. A dedicated derivation for Gaussian smoothing is beyond the scope of the present work. Nevertheless, we use the above formalism as a useful approximation to investigate the impact of non-linear corrections with the Gaussian window function in Appendix~\ref{appsec:constraints-tophat}.

\subsection{Effect of non-sphericity of the overdense region}

The overdense region in more realistic scenarios is not perfectly spherical; therefore, to precisely estimate the threshold density contrast, the non-sphericity should be taken into account.
We model the non-spherical overdense region as an ellipsoid and account for non-spherical effects up to first order in ellipticity. 
	As discussed in Appendix \ref{appendix-nr}, 
	there is no correction to the non-linear relation \eqref{delta-rho-nl-flrw} to this order.
	Meanwhile, the threshold density contrast is given by $\delta_{\rm ec} \simeq \delta_{ m,c}( 1+ 3e )$, 
	where $e \simeq \frac{3\sigma_L}{\sqrt{10\pi}\delta_{L,c}}$ is the ellipticity 
	and $\delta_{m,c} $ is the threshold for the spherical overdensity~\cite{Kuhnel:2016exn,2018-Sasaki.etal-CQG,Escriva:2024lmm,Escriva:2024aeo}. 
This relation suggests that a larger threshold density contrast is needed for ellipsoidal overdense regions to form PBHs as compared to spherical shapes. Incorporating this non-spherical threshold into the relation between the non-linear density contrast $\delta_m$ and the linear density contrast $\delta_L$ yields
	\begin{align}\label{deltam-deltal-relation-el}
		\delta_{m,c} \left( 1+ \frac{9\sigma_L}{\sqrt{10\pi}\delta_{L,c} } \right)  =  \delta_{L,c} - \frac{3}{8}  \delta_{L,c}^2, ~~
	\end{align}
	This is a cubic equation for $\delta_{L,c}$. Since we work to first order in $e$, and solving this cubic equation for $\delta_L$ up to first order in the ellipticity, the linear threshold for type-I PBH formation becomes
	\begin{align}\label{delta-ec-new}
		\delta_{L, \rm ec} \simeq \delta_{L-,c} + \frac{ (27/\sqrt{10\pi}) \delta_{m,c} \sigma_L}{6 \delta_{m,c} - 4 + 4\sqrt{1-\frac{3}{2} \delta_{m,c}} }~~.
	\end{align}
	This relation incorporates both non-linear effects and non-sphericity into the threshold for the linear density contrast $\delta_L$. We use $\delta_{L,\rm ec}$ in place of $\delta_{L-,c}$ as the lower limit of
	the integral in the PBH abundance when the ellipsoidal collapse is considered.

\section{PBH mass function: A brief review}\label{sec-pbh-computation}%

In this section, we provide a brief overview of computing the PBH mass function from the primordial curvature perturbation, following earlier works~\cite{2018-Sasaki.etal-CQG}. 
Throughout this work we restrict ourselves to Gaussian curvature perturbations, and neglect the effects of primordial non-Gaussianity.

\subsection{Definition of the PBH mass function}
We define the PBH mass function $\psi (M)$ such that $\psi (M) dM$ represents the 
mass fraction of PBHs whose masses are in the infinitesimal mass range $(M,M+dM)$ with
the normalization factor set by total dark matter abundance
\begin{align}\label{app-fpbh-def}
	\int  d M \, \psi (M)= f_{\rm PBH}.
\end{align}
Here $f_{\rm PBH} \equiv \frac{\Omega_{\rm PBH}}{\Omega_{\rm CDM}}$ is the fraction
of PBHs in dark matter and $\Omega_{\rm CDM}=0.265$ is the matter density parameter at the present epoch.
We emphasize that this definition of the PBH mass function is independent of the PBH formation 
mechanism.

\subsection{PBH mass function and the mass fraction parameter $\beta$}
To constrain the primordial power spectrum from the abundance of PBHs, we need to obtain the relation between the PBH mass function and primordial power spectrum. However, this relation is inherently ambiguous, since no proposed method provides a rigorous first-principles derivation of the PBH mass function. 
As a result, one must rely on phenomenological approaches motivated by physical intuition, leading to several approximate and mutually incomplete expressions for the mass function.
In this paper, we adopt the postulate given in the literature \cite{Carr:2009jm} that the so-called mass fraction parameter 
usually denoted by $\beta (M)$, whose concrete expression will be given later, 
represents the mass fraction of PBHs evaluated
at their formation time in the logarithmic mass bin:
\begin{equation}
	\Omega_{\rm PBH}=\int \beta (M) {\left( \frac{M_{\rm eq}}{M} \right)}^\frac{1}{2} d\ln M.
\end{equation}
Here $M_{\rm eq}$ is the horizon mass at the time of the
matter-radiation equality and 
the factor ${\left( \frac{M_{\rm eq}}{M} \right)}^\frac{1}{2}$ accounts for the fact
that PBH abundance relative to radiation grows in proportional to the scale factor 
during the RD era.
Comparing this relation with the definition of the mass function (\ref{app-fpbh-def}) immediately yields
\begin{align}\label{app-Omega-mpbh}
	\psi (M_{\rm PBH})= - \frac{2}{\Omega_{\rm CDM} k_{\rm eq}} 
	\frac{dk_m}{d M_{\rm PBH}}\beta (k_m)~,
\end{align}
where $k_{\rm eq}$ is the comoving wavenumber of perturbations that reenter the Hubble
horizon at the time of the matter-radiation equality and 
$k_m=1/r_m$ is related to the PBH mass $M_{\rm PBH}$ by $k_m =a_m/M_{\rm PBH}$
($a_m$ is the value of the scale factor evaluated at the time when the scale $r_m$ reenters
the Hubble horizon).
Notice that $dk_m/dM_{\rm PBH}$ is negative.
In this framework, the computation of the PBH mass function boils down to the 
computation of $\beta (k_m)$, following the relation \footnote{Alternatively, we can also define $\psi (M)$ as dimensionless quantity by $\psi (M) = \frac{1}{\Omega_{\rm CDM}} \frac{d \Omega_{\rm PBH}}{d \ln{M}}$, such that $f_{\rm PBH} =\int  d \ln{M} \, \psi (M)$, see \cite{Gow:2020bzo,Young:2024jsu}. One can also define the mass function as the derivative of the probability (i.e., $\beta(M)$), however, as it is shown in our previous work~\cite{Kushwaha:2025zpz}, this may lead to the negative PBH mass function (thus, unphysical) in certain mass ranges. Since no method currently provides a rigorous first-principles derivation of the mass function. As a result, one should rely on phenomenological approaches motivated by physical intuition, yielding several approximate, mutually incomplete expressions for the mass function.}
\begin{align}\label{psi-mpbh-rel-fin}
	\psi (M_{\rm PBH})= \frac{1}{\Omega_{\rm CDM}} 
	\left( \frac{g_{*}}{10.75} \right)^{-1/12} \left( \frac{2.9\times 10^5 \, \rm{Mpc}^{-1}}{k_{\rm eq}} \right)\left(\frac{30 M_{\odot}}{M_{\rm PBH}} \right)^{1/2} \left(\frac{1}{M_{\rm PBH}}\right) \, \beta (M_{\rm PBH} (k_m) )~~.
\end{align}
In the literature, there are two representative approaches that connect $\beta$
to the primordial power spectrum.
They are PS formalism and the peak theory\footnote{Extended formulations of peak theory and the excursion-set approach have also been developed to describe PBH formation, see Refs.~\cite{Yoo:2018kvb,Germani:2019zez,Young:2020xmk,Kushwaha:2025zpz}.}. In what follows,
we review how $\beta$ is related to ${\cal P}_\zeta$ in each formalism.

\subsection{Mass fraction parameter in PS formalism}
In the PS formalism, the mass fraction parameter $\beta$ is defined as~\cite{1974-Press-Schechter-APJ}
\begin{align}\label{beta-ps}
	\beta (k_m) = \int_{\delta_{L-,c}}^{\delta_{\rm max}} d\delta_L ~ \frac{M_{\rm PBH}(\delta_L)}{M_H} P (\delta_L), 
\end{align}
where $\delta_L$ is the density contrast smoothed over the scale $r_m=1/k_m$,
$\delta_{\rm max}=4/3$ is the maximum value of $\delta_L$ corresponding to the maxima $C_{L,\rm max}$ and $P (\delta_L)$ is the probability distribution function (PDF) of $\delta_L$.
The PBH mass $M_{\rm PBH}$ depends on $\delta_L$ when the critical phenomena is taken into account (see \ref{subsec:critialp})
and $M_H$ is the horizon mass when the scale $r_m=1/k_m$ reenters the Hubble horizon.
Often, a factor of `2' is multiplied in the above formula following the original PS formalism~\cite{1974-Press-Schechter-APJ}, which is related to the `cloud-in-cloud' problem. 
However, in our previous work~\cite{Kushwaha:2025zpz}, using the excursion set formalism, 
we have shown that the prescription of multiplying by the factor `2' is not justified
for computing the abundance of PBHs formed during the RD era. 
In this paper, we assume that the multiplication factor is one and adopt Eq.~(\ref{beta-ps})
to compute the PBH abundance.
Although, for the purpose of constraining the primordial power spectrum from PBHs abundance, this multiplicative factor would approximately give the same results.

Because $\zeta$ is assumed to follow Gaussian statistics, $\delta_L$ is also Gaussian. 
Thus, the PDF of $\delta_L$ is given by
\begin{align}\label{pdf-del-gaussian}
	P (\delta_L) = \frac{1}{\sqrt{2\pi} \, \sigma_L} \exp{\left(-\frac{\delta_L^2}{2\sigma_L^2}\right)} ~,
\end{align}
where $\sigma_L^2 = \langle \delta_L^2 \rangle $ is the variance of the density contrast $\delta_L$ and from Eq.~(\ref{relation-deltaL-zeta}), it is related to the primordial spectrum of $\zeta$ by
\begin{align}\label{sigma-l}
	\sigma_L^2= \frac{16}{81} \int_0^{\infty} \frac{dk}{k} \, (kr_m)^4 \tilde{W}^2(k,r_m) T^2 (k,r_m) \mathcal{P}_{\zeta} (k) ~~,
\end{align}
The PBH mass fraction is then given by
\begin{align}\label{beta-ps-final}
	\beta (k_m)= \int_{\delta_{L-,c}}^{\delta_{\rm max}} d\delta_L ~ \left( \frac{M_{\rm PBH} (\delta_L)}{M_H} \right) \frac{1}{\sqrt{2\pi} \sigma_L} \exp{\left(-\frac{\delta_L^2}{2\sigma_L^2}\right)} ~~.
\end{align}
As we can see from the above equation, the PBH mass fraction is related to the primordial curvature power spectrum $\mathcal{P}_{\zeta} (k)$ via the variance $\sigma_L^2$ in Eq.\eqref{sigma-l}. 
Therefore, given the constraints on PBHs' abundance, we can invert the above equation to constraints on ${\cal P}_{\zeta} (k)$.

\subsection{Mass fraction parameter in the peak theory}
The peak theory is based on the assumption that the peaks in smoothed linear density field $\delta_L$ correspond to peaks in smoothed nonlinear density field such that $\delta_m = C_{\rm NL}$, where $C_{\rm NL}$ in Eq.\eqref{cnl-final} can be applied to calculate the height of the peaks in nonlinear field (although this statement may not be true in general, but will be valid in the case that only sufficiently rare and large peaks are considered, see Ref.~\cite{Young:2019yug}).
The mass fraction parameter $\beta$ is related to the number density of the peaks as
\begin{align}\label{beta-peak-theory-def}
	\beta (k_m) = \frac{(2\pi)^{3/2} }{(a H)^3} \int_{\delta_{L-,c}}^{\delta_{\rm max}} d\delta_L ~~ \frac{M_{\rm PBH}(\delta_L)}{M_H} \, n\left(\frac{\delta_L}{\sigma_L} \right),
\end{align}
where $(aH)^{-1}$ is the horizon size at the re-entry.
The number density $n\left(\frac{\delta_L}{\sigma_L} \right)$ for a random Gaussian field of rare peaks whose density contrast takes a value in $(\delta_L,\delta_L+d\delta_L)$ in a comoving volume is given by~\cite{Gow:2020bzo,Young:2014ana}
\begin{align}\label{n-peaks}
	n\left( \frac{\delta_L}{\sigma_L} \right) = \frac{\mu_L^3}{3^{3/2} 4\pi^2 \sigma_L^3} \left( \frac{\delta_L}{\sigma_L} \right)^3  e^{-\frac{\delta_L^2}{2\sigma_L^2}}~~,
\end{align}
where $\mu_L$ is given by
\begin{align}\label{mu-l}
	\mu_L^2 = \int_0^{\infty}  \frac{dk}{k} \, k^2 \mathcal{P}_{\delta_L} (k) = \frac{16}{81} \int_0^{\infty} \frac{dk}{k} \, (kr_m)^4 \tilde{W}^2(k,r_m) T^2 (k,r_m)   \, k^2\mathcal{P}_{\zeta} (k) ~~.
\end{align}
Using Eq.\eqref{n-peaks} in Eq.\eqref{beta-peak-theory-def}, we obtain
\begin{align}\label{beta-peak-theory-final}
	\beta (k_m) = \frac{1}{3^{3/2}  \sqrt{ 2\pi} (a H)^3} \int_{\delta_{L-,c}}^{\delta_{\rm max}} d\delta_L ~ \left( \frac{M_{\rm PBH}(\delta_L)}{M_H} \right) \, \left( \frac{\mu_L}{\sigma_L}\right)^3 \left( \frac{\delta_L}{\sigma_L} \right)^3  e^{-\frac{\delta_L^2}{2\sigma_L^2}} ~,
\end{align}
where we used $\delta_c = \delta_{L-,c}$ and $(aH)^{-3}=r_m^3$.

\subsection{PBH mass in the critical phenomena}
\label{subsec:critialp}
Before closing this section, let us briefly discuss the relation between $M_{\rm PBH}$ and the horizon mass $M_H$ (which is proportional to the square of the horizon scale, $(aH)^{-2} = r_m^2$), which appears in the estimation of mass fraction parameter in Eq.~\eqref{beta-ps-final} and Eq.~\eqref{beta-peak-theory-final}. The final mass of the PBHs formed during the RD era, $M_{\rm PBH}$ depends on the initial density perturbation profile (i.e., shape and amplitude) and follows the scaling behaviour~\cite{1998-Niemeyer.Jedamzik-PRL}
\begin{align}\label{mpbh-mh-scaling}
	M_{\rm PBH} = \kappa M_H (\delta - \delta_c)^{\gamma} = \kappa M_H \left[ \delta_L - \frac{3}{8} \delta_L^2 - \delta_c \right]^{\gamma}
\end{align}
where $\kappa$ (which we take 3 in this work) depends on the shape of the profile, $\gamma \simeq 0.36$  during RD epoch~\cite{Evans:1994pj,Young:2019yug}. 
This scaling behaviour holds approximately true up to $M_H$.
The near-critical behaviour~\eqref{mpbh-mh-scaling} shows that even if we consider perturbations of a single mode leading to PBHs formation, the PBHs are formed in a finite mass range;
and the monochromatic mass function, (defined here as a final PBH mass distribution in which all PBHs have the same mass) is never realized.
Nevertheless, the spread of the mass function is not significant and we may practically treat
the resultant mass function as monochromatic,
which only makes a subdominant contribution to the PBH abundance.
In what follows, we also consider the monochromatic mass function by assuming that all the 
PBHs have the same mass parametrized by a parameter $\alpha$ defined by $\alpha=M_{\rm PBH}/M_H $, where $M_H= \frac{4\pi }{3} \rho H^{-3}$.

Lastly, we should also note that when considering the constraints on monochromatic PBH mass function that are produced by a very narrow peaked power spectrum that peaks at a given wavenumber $k_p$, we can identify\footnote{The identification $r_m = 1/k_p$ is a simplifying approximation. For instance, for a sinc-shaped typical profile, maximizing the compaction function gives $r_m \simeq 2.74/k_p$~\cite{Pi:2024ert}, and the ratio becomes profile dependent for broader spectra~\cite{Pi:2024ert}. Adopting this refined identification would shift the constraints on $k_p$ and might affect the inferred amplitude.} $r_m = 1/k_p$. This gives the PBHs mass fraction as a function of peak wavenumber, $\beta (k_p)$. Furthermore, we can relate this theoretical estimate with the constraints by using the mass-wavenumber relation~\cite{2018-Sasaki.etal-CQG}
\begin{align}\label{mpbh-k-relation}
	M_{\rm PBH} (k) \simeq 30 M_{\odot} \left( \frac{\alpha}{0.2} \right) \left( \frac{g_{*}}{10.75} \right)^{-1/6} \left(\frac{k}{2.9\times 10^5 \, \rm{Mpc}^{-1}} \right)^{-2} ~,
\end{align}
$g_{*}$ is the number of relativistic degrees of freedom at the epoch of PBH formation.

\section{Constraints on $\mathcal{P}_\zeta (k)$ from PBHs abundance}
\label{sec-constraint}
In this section, we derive model-independent (i.e., independent of the production mechanism of primordial perturbations, under the assumption of Gaussian statistics) constraints on the primordial curvature power spectrum by using the PBH mass function derived in the previous section. 
We consider a generic form of the lognormal-type power spectrum, which can be parametrized as~\cite{Pi:2020otn}
\begin{align}\label{lognormal-powerspect}
	\mathcal{P}_{\zeta} (k) = \mathcal{A}_p \frac{1}{\sqrt{2\pi} \Delta} \exp{\left( - \frac{\ln{(k/k_p)}^2}{2\Delta^2}\right)}~~,
\end{align}
where $k_p$ and $\Delta$ are free parameters determining the position and the width of the peak, respectively. The overall normalization is determined by the amplitude $\mathcal{A}_p$ which satisfies the condition $\int_0^{\infty} \mathcal{P}_{\zeta} (k) d\ln{k} = \mathcal{A}_p$.
Note that the lognormal peak in the primordial curvature power spectrum can arise in many inflation models, for example, models predicting a very narrow peak in the primordial curvature power spectrum with $\Delta \ll 1$ are discussed in Refs.\cite{Kawasaki:1997ju,Kawasaki:2012wr,Inomata:2017okj}, and the models supporting the broad power spectrum with $\Delta \geq \mathcal{O}(1)$ are discussed in Refs.\cite{Garcia-Bellido:1996mdl,Yokoyama:1998pt,Clesse:2015wea,Garcia-Bellido:2017mdw,Inomata:2018cht,Braglia:2020eai}.
Interestingly, it is shown in Ref.\cite{Gow:2020bzo} that the PBH mass function remains approximately the same for $\Delta \lesssim 0.1$. Thus, we can use the lognormal power spectrum \eqref{lognormal-powerspect} with $\Delta = 0.1$ as the typical case for a monochromatic mass function of PBH (accordingly, in Figure.\ref{fig:Ap-Pzeta-PS-PT-near-critical}, we approximate the resulting PBH mass function as monochromatic).

\subsection{Constraints on narrow peaked power spectrum from monochromatic PBH mass function}\label{subsec-monochromatic}

The narrow-peaked power spectrum produces all the PBHs with the same mass, which are described by a monochromatic mass function of PBHs.
The constraints on PBHs' abundance for the monochromatic PBH mass function are presented in terms of the rescaled parameter $\beta'$, which is defined as~\cite{2021-Carr.etal-Rept.Prog.Phys}
\begin{align}\label{app-beta-prime-def}
	\beta^{\prime} (M) = \alpha^{1/2} \left( \frac{g_{*,i}}{106.75} \right)^{-1/4} \left(\frac{h}{0.67} \right)^{-2} \beta (M)
\end{align}
where $h$ is the uncertainty in Hubble parameter and $g_{*,i}$ is the number of relativistic degrees of freedom at the epoch of PBH formation\footnote{The advantage of using $\beta^{\prime} (M) \approx 7.06 \times \Omega_{\rm PBH} (M) \left( \frac{M}{10^{15} \, \rm g}\right)^{1/2}$ instead of $\beta$ as observational quantity is that it is independent of uncertainty in the values of $\alpha, h, g_{*,i}$, where $\Omega_{\rm PBH} \equiv \rho_{\rm PBH} (t_0) / \rho_{\rm critical}$ is the PBH density parameter at present epoch.}.
\begin{figure*}[t!]
	\centering
	\includegraphics[height=2.6in,width=6.4in]{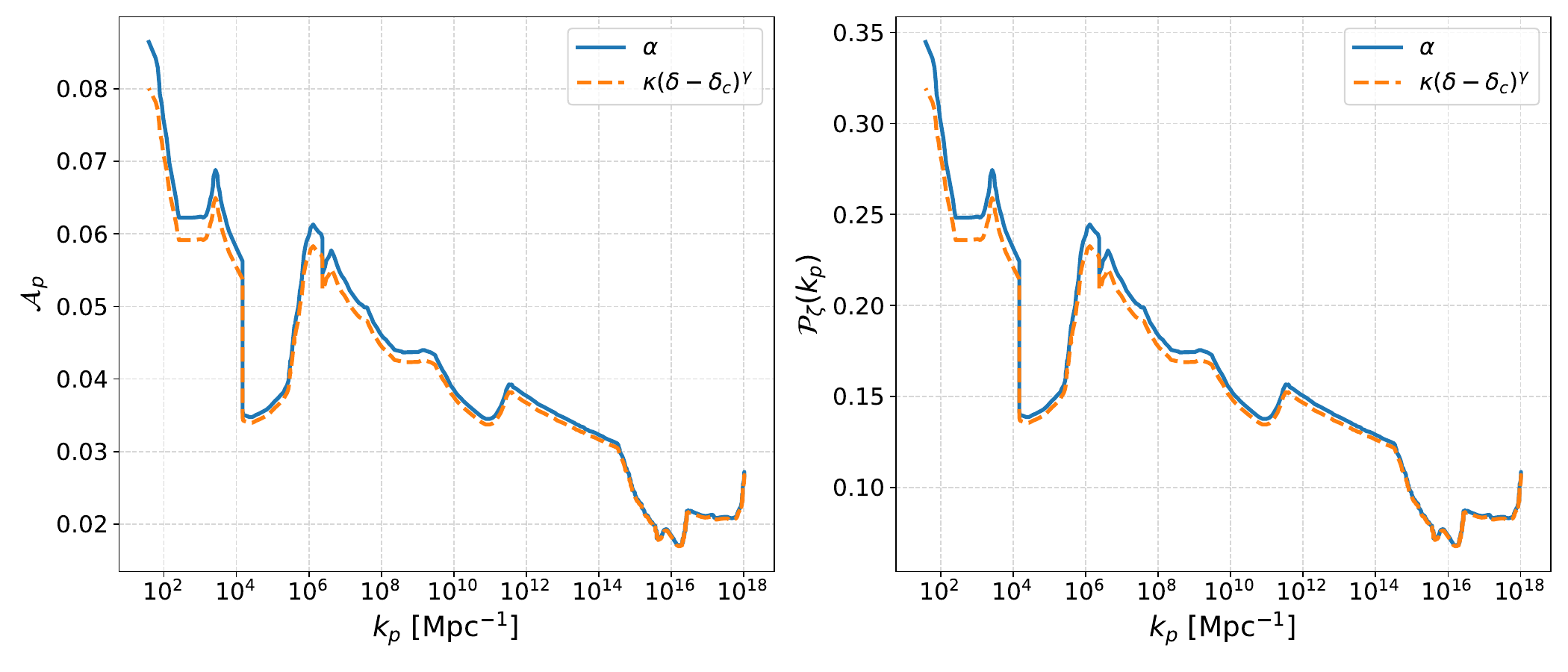} 
\includegraphics[height=2.6in,width=6.4in]{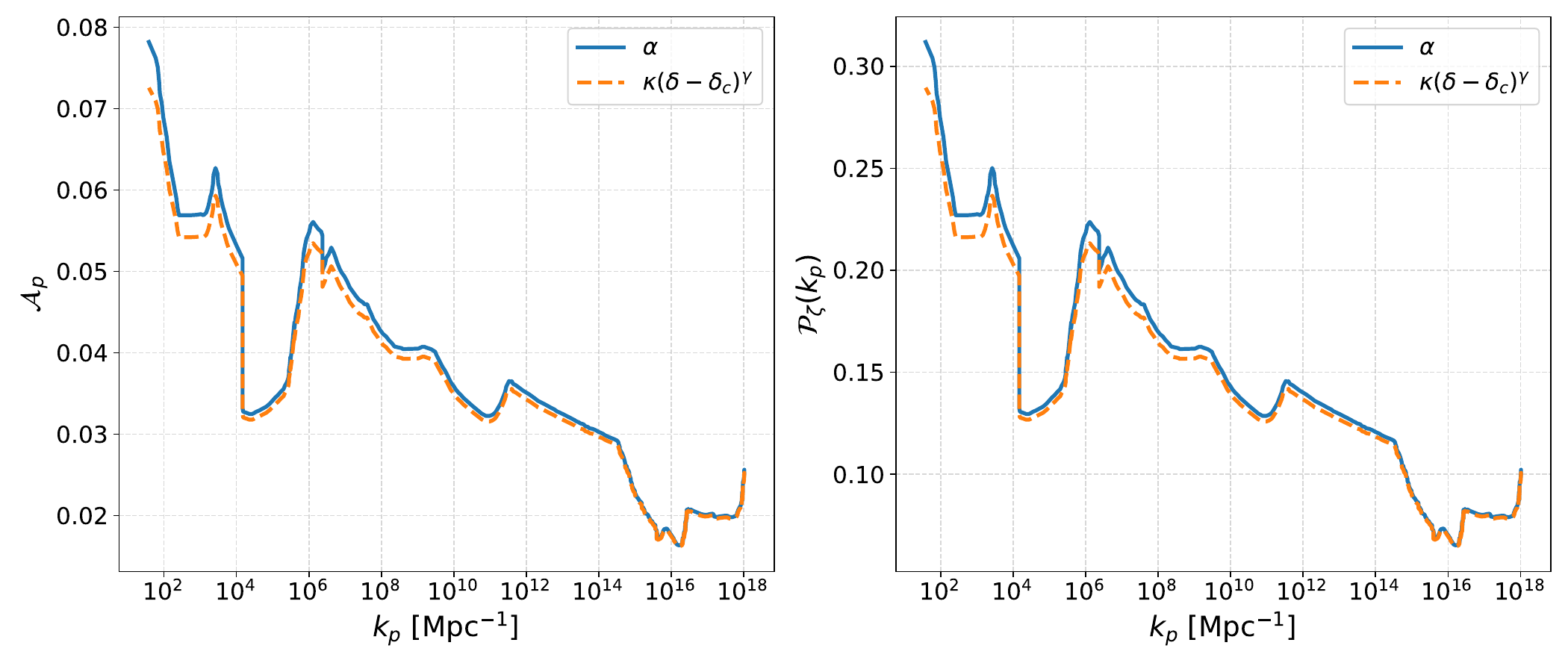} 
\caption{The effect of the mass-ratio $M_{\rm PBH}/M_H$ on the amplitude $\mathcal{A}_p$ and the lognormal power spectrum $\mathcal{P}_{\zeta} (k_p)$ for $\Delta=0.1$ for spherical overdensity case. Solid curves are for constant value $\alpha=0.2$ and dashed curves show the ratio obtained from the near-critical behaviour~\eqref{mpbh-mh-scaling}. The upper panel is for the PS formalism, and the lower panel is for the peak theory. In generating these plots, we choose $\alpha=0.2$ in relation between $\beta$ and $\beta'$ in Eq.\eqref{app-beta-prime-def} in this work, and near-critical behaviour is assumed only inside the integration in the expression of $\beta$ in Eq.\eqref{beta-ps-final} and Eq.\eqref{beta-peak-theory-final}. These plots are for the case of real-space top-hat window function with transfer function, and we take $\kappa=3, \gamma =0.3$ with $\delta_c =0.45$}.
\label{fig:Ap-Pzeta-PS-PT-near-critical}
\end{figure*}
Strictly speaking, the monochromatic mass function corresponds to the case where the mass-ratio, $\alpha $ is constant. Although $\alpha$ depends on the details of the gravitational collapse, we take $\alpha \simeq 0.2$, see the Ref.~\cite{2018-Sasaki.etal-CQG}. 
Another relation for the mass-ratio is determined by using the near-critical behavior in Eq.\eqref{mpbh-mh-scaling}, which gives $\frac{M_{\rm PBH}}{M_H} = \kappa (\delta - \delta_c)^{\gamma}$. Figure~\ref{fig:Ap-Pzeta-PS-PT-near-critical} compares the constraints on the primordial power spectrum obtained using the near-critical scaling relation with those derived assuming a constant mass ratio, $\alpha=0.2$, for spherical collapse and a narrow power-spectrum peak ($\Delta=0.1$).
We find that the inclusion of near-critical collapse leads to only minor differences in the inferred power-spectrum amplitude. We have also verified that a similar conclusion holds for a broader peak ($\Delta=1$). Consequently, for the parameter range considered in this work, adopting the near-critical scaling relation yields constraints that are very similar to those obtained using the commonly assumed constant mass ratio $\alpha$.
Therefore, we show our final constraints from the monochromatic mass function for the case of $\alpha=0.2$ only.
\begin{figure*}[t!]
\centering
\includegraphics[height=2.6in,width=6.4in]{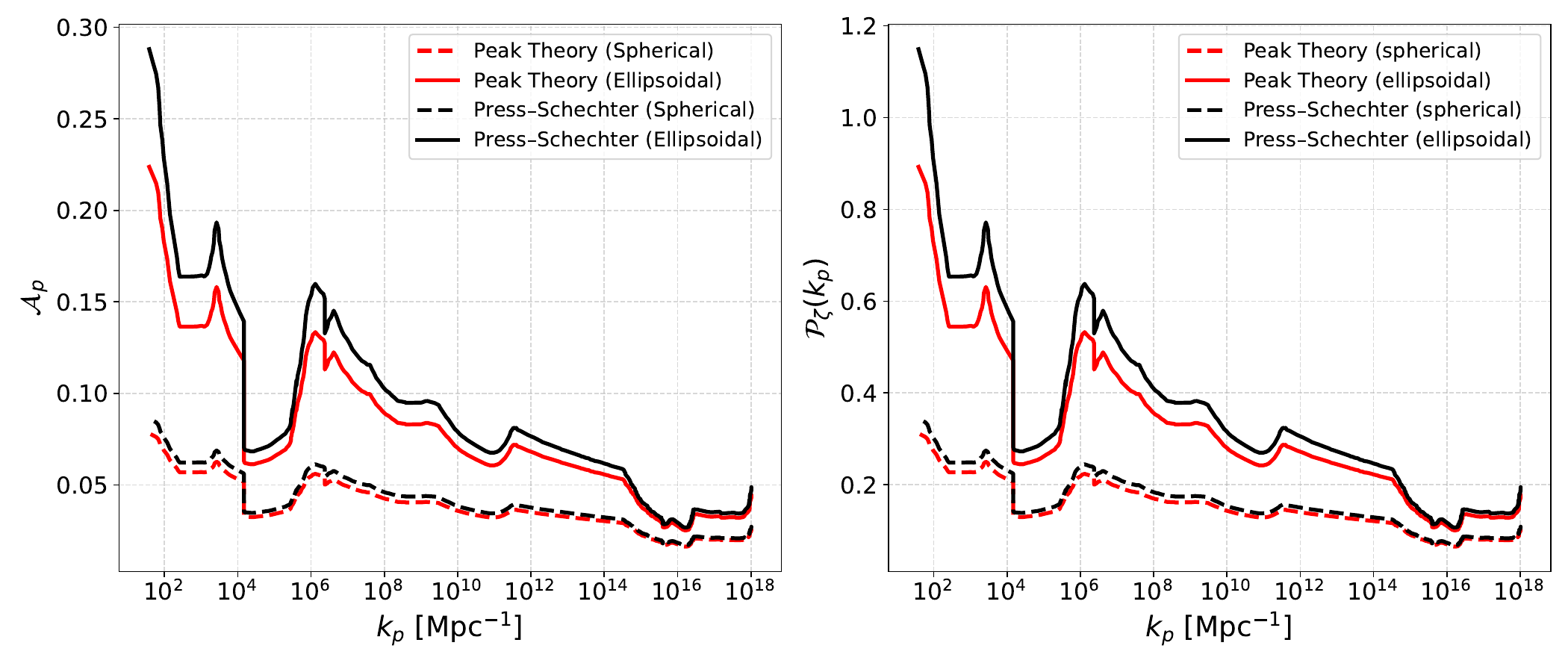} 
\caption{Comparison between Press-Schechter and peak theory on the constraints on $\mathcal{A}_p$ and $\mathcal{P}_{\zeta} (k_p)$ for both spherical and ellipsoidal overdensity case, we use $\Delta=0.1$ and $\alpha=0.2$.}
\label{fig:Ap-Pzeta-mc-PS-pt}
\end{figure*}
To obtain the constraints on the primordial curvature power spectrum,
we use $\beta(k_p)$ derived in Eq.~\eqref{beta-ps-final} within the PS formalism and in Eq.~\eqref{beta-peak-theory-final} within peak theory, and substitute these into the rescaled PBH mass fraction $\beta'(M)$
defined in Eq.\eqref{app-beta-prime-def}. 
These expressions follow from the refined calculations based on the compaction function and incorporate the non-linear relation between the density contrast and the curvature perturbation.
Using the most up-to-date constraints on the PBHs abundance from Ref.~\cite{2021-Carr.etal-Rept.Prog.Phys} together with the relation between PBHs mass and peak wavenumber in Eq.\eqref{mpbh-k-relation}, we derive the corresponding constraints on the amplitude of the primordial curvature power spectrum assuming a monochromatic mass function. The results are shown in Figure~\ref{fig:Ap-Pzeta-mc-PS-pt}, which compares the constraints obtained using the PS formalism and peak theory for both spherical and non-spherical (ellipsoidal) overdense regions. 
As expected, since the threshold density contrast for collapse is larger in the ellipsoidal case than in the spherical case, a larger amplitude of the power spectrum is required to produce the same PBH abundance. Some of these effects have been studied previously in the literature. For example, the comparison between the PS formalism and peak theory was investigated in Refs.~\cite{Kalaja:2019uju,Gow:2020bzo}, while Ref.~\cite{Sato-Polito:2019hws} considered the impact of non-sphericity within the PS formalism. However, these studies employed older PBH abundance constraints. Therefore, Fig.~\ref{fig:Ap-Pzeta-mc-PS-pt} presents updated and more robust constraints by comparing all of the above mentioned effects.
\begin{figure*}[t!]
\centering
\includegraphics[height=3.8in,width=6in]{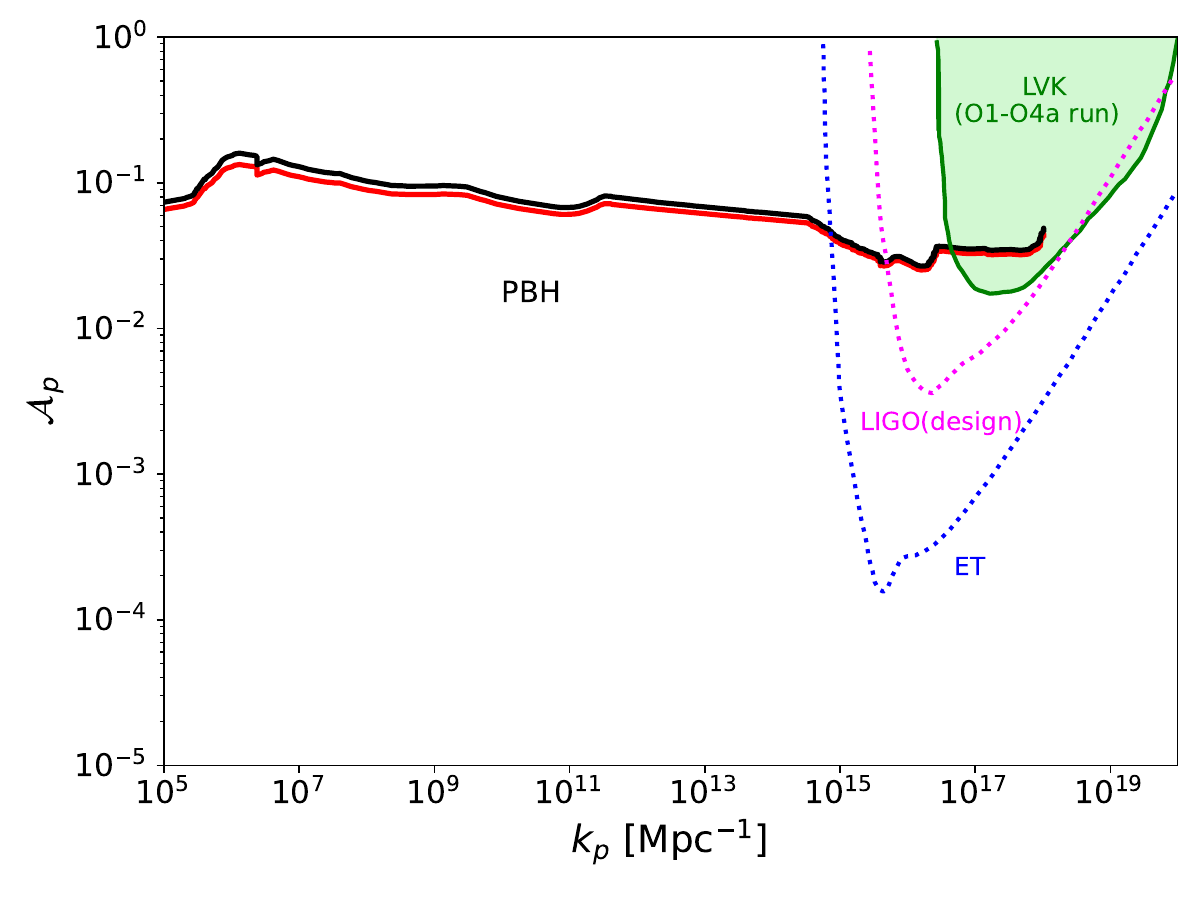} 
\caption{Comparison of our constraints for a narrow power spectrum (ellipsoidal case) from Fig.~\ref{fig:Ap-Pzeta-mc-PS-pt} with other observational limits. The red and black curves show the constraints for peak theory and PS formalism, respectively.
The solid Green line shows the region constrained by LVK $O1-O4a$ run~\cite{LIGOScientific:2025kry}. The dotted magenta and blue curves show the
projected sensitivities of the LIGO in its final phase (LIGO design) and the Einstein Telescope (ET), curves are taken from~\cite{Romero-Rodriguez:2021aws}. }
\label{fig:Ap-constraint-comp}
\end{figure*}
Furthermore, in Figure~\ref{fig:Ap-constraint-comp}, we compare our constraints (for the ellipoidal case) on the amplitude of primordial power spectrum $\mathcal{A}_p$ with the recent constraints from the LIGO-Virgo-KAGRA (LVK) O1-O4a runs~\cite{LIGOScientific:2025kry}, and the projected sensitivities of the Einstein Telescope (ET) and LIGO design (i.e., LIGO in its final phase).

\subsection{Constraints on broad peaked power spectrum from extended PBHs mass function}
\label{sec-extended-mass-constraints}

\begin{figure*}[t!]
\centering
\includegraphics[height=2.6in,width=6.4in]{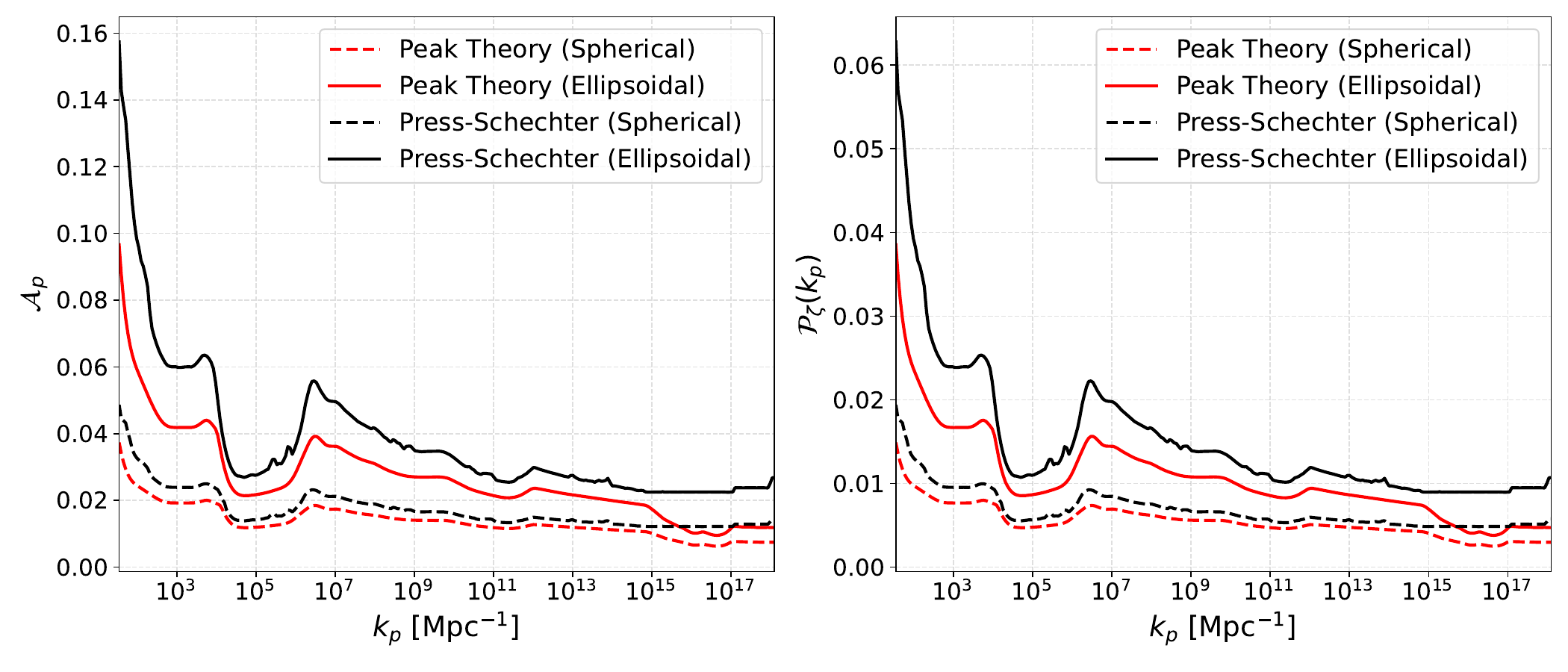} 
\caption{Comparison between Press-Schechter and peak theory on the constraints on $\mathcal{A}_p$ and $\mathcal{P}_{\zeta} (k_p)$ for both spherical and ellipsoidal overdensity case, for broad peak case where we use $\Delta=1$, $\alpha=0.2$ and $\delta_c = 0.45$. The constraints are obtained from the PBHs abundance against DM, $f_{PBH}$. 
The oscillatory behaviour for large $k_p$ are the numerical artifact of rapidly oscillating $j_1^2 (k r_m) $ term in $\sigma_L^2$ and $\mu^2_L$ in Eqns.\eqref{sigma-l} and \eqref{mu-l}. The oscillating features are insignificant for the peak theory due to the effective cancellation between both terms in Eq.\eqref{beta-peak-theory-final}, while they become more prominant for PS formalism due to $\sigma_L^2$ dependence in Eq.\eqref{beta-ps-final}. Consequently, the apparent flattening of the Press-Schechter constraints in the range $10^{15}<k_p<10^{17}$ is likely not physical but a numerical artifact. The underlying behaviour is expected to qualitatively follow that of the peak-theory constraints.}
\label{appfig:Ap-Pzeta-mc-PS-pt}
\end{figure*}

\begin{figure*}[t!]
\centering
\includegraphics[height=4in,width=6in]{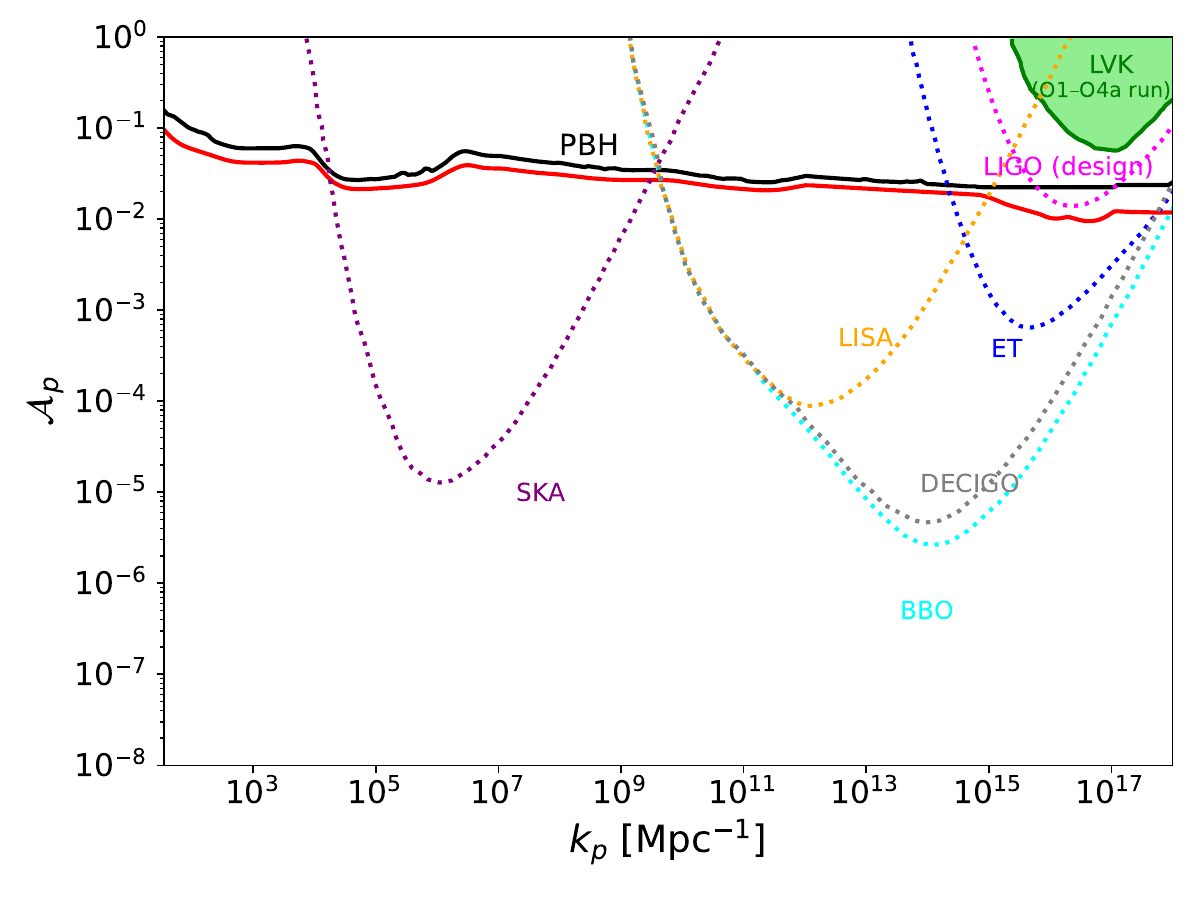} 
\caption{Comparison of our constraints for a broad power spectrum (ellipsoidal case) from Fig.~\ref{appfig:Ap-Pzeta-mc-PS-pt} with other observational limits. The red and black curves show the constraints for peak theory and PS formalism, respectively.
The solid Green line shows the region constrained by LVK $O1-O4a$ run~\cite{LIGOScientific:2025kry}. The dotted curves show the
projected sensitivities of the LIGO in its final phase (LIGO design) and the ET, SKA, BBO, DECIGO, LISA, which are taken from~\cite{Romero-Rodriguez:2021aws,Inomata:2018epa}. 
}
\label{ap-fig:Ap-constraint-comp-broad}
\end{figure*}

Using the PBHs mass function Eq.\eqref{psi-mpbh-rel-fin}, 
we derive the constraints on the broad peak primordial power spectrum $\mathcal{P}_{\zeta}(k_p)$ for $\Delta=1$, using the method proposed in \cite{Carr:2017jsz} (also summarised in appendix \ref{sec-extended-mass} for completeness). The constraints\footnote{The constraints in Figures~\ref{appfig:Ap-Pzeta-mc-PS-pt} and \ref{ap-fig:Ap-constraint-comp-broad} are derived from the PBH abundance relative to dark matter $f_{\rm PBH}$, which is related to the initial abundance $\beta'(M)$ through 
$f_{\rm PBH}=2.7\times10^8 (M/M_\odot)^{-1/2}\beta'(M)$.} for both PS and peak theory are shown in Figure \ref{appfig:Ap-Pzeta-mc-PS-pt}. Similar to the monochromatic case, we can see that the amplitude of the primordial power spectrum is larger for the ellipsoidal case. %
In Figure.~\ref{ap-fig:Ap-constraint-comp-broad}, we compare our constraints with other observations probing small-scale primordial power spectrum, such as LIGO-Virgo-KAGRA (LVK) O1-O4a runs~\cite{LIGOScientific:2025kry}, ET~\cite{2009-Sathyaprakash.Schutz-LivRevRel,Moore:2014lga}, Square Kilometer Array (SKA)~\cite{Janssen:2014dka,Moore:2014lga}, Big Bang Observer (BBO)~\cite{Yagi:2011wg,Thrane:2013oya}, DECi-hertz Interferometer Gravitational-wave Observatory (DECIGO)~\cite{Kuroyanagi:2014qza}, Laser Interferometer Space Antenna (LISA)~\cite{Thrane:2013oya,2017arXiv170200786A}. 
Furthremore, we also present our constraints for the broad power spectra for Gaussian window function in  appendix~\ref{appsec:constraints-tophat} (see Figures~\ref{appfig:Ap-Pzeta-PS-pt-gausswindow-broad} and \ref{ap-fig:Ap-constraint-comp-broad-gausswindow}).

\section{Conclusion}
\label{sec-conclusion}
Constraints on the abundance of PBHs provide a unique opportunity to constrain the primordial power spectrum ($\mathcal{P}_\zeta$) at small scales, regimes which are not probed by standard cosmological observations like the CMB and galaxy surveys. This connection is established via the PBH formation mechanism, which relies on the collapse of large-amplitude primordial curvature perturbations at the time of the horizon re-entry during RD epoch. Crucially, deriving stringent limits requires the precise and careful modeling of the nonlinear dynamics of the collapse of density perturbations leading to PBH formation in the early Universe.

In this work, we revisited and refined the methodology to translate the PBH abundance bounds into upper-limits on the primordial curvature power spectrum, thereby placing this connection on a more solid theoretical footing. Using the most up-to-date constraints on PBH abundances, we systematically investigated multiple sources of theoretical uncertainty that affect PBH formation and their resulting mass function. Employing refined calculations based on the compaction function and accounting for the full nonlinear relation between curvature perturbations and density contrast, we compared the resulting constraints across key modeling choices, including the PS formalism versus peak theory and monochromatic versus extended PBH mass functions, different choices of window functions and the inclusion of the transfer function, while also assessing the impact of spherical versus ellipsoidal collapse criteria for overdense regions.
Although several of these effects have been examined individually in earlier works, to our knowledge, this is the first detailed study that incorporates all of them within a single, unified framework to evaluate their combined impact on the inferred primordial power spectrum. Our results showed that the amplitude of primordial power spectrum (for both narrow and broad peak) required for the PBHs formation is larger when non-spherical collapse is included, owing to the higher threshold density contrast.
By incorporating various effects coherently, our analysis provides more robust, precise upper limits on both narrow and broad peak primordial curvature power spectra. We emphasize that these limits assume Gaussian primordial curvature perturbations; since local-type non-Gaussianity can substantially modify the probability of rare, large fluctuations, the inferred constraints may change significantly in non-Gaussian scenarios, which we leave for future work. These results not only refine the constraints on PBH scenarios but also strengthen the use of PBH abundance as an indirect probe of the final stages of inflation, particularly on smaller scales inaccessible to other cosmological observations. In Figures \ref{fig:Ap-constraint-comp} and \ref{ap-fig:Ap-constraint-comp-broad}, we also compared our constraints with other observations. We also investigated the impact of the window and transfer functions on the inferred constraints. While the transfer function produces only modest corrections when a Gaussian window function is employed, its effect can become significant for the real-space top-hat window function, particularly in the case of broad primordial power spectra, where neglecting the transfer function can lead to significant discrepancies in the inferred constraints at small scales. These results highlight an additional source of theoretical uncertainty in translating PBH abundance constraints into limits on the primordial curvature power spectrum.

\acknowledgments
We are grateful to the referee for insightful comments and suggestions that improved the quality and clarity of this work.
The work of A.K. was supported by the Japan Society for the Promotion of Science (JSPS) as part of the JSPS Postdoctoral Program (Grant Number: 25KF0107). 
T.S. gratefully acknowledges support from JSPS KAKENHI grant (Grant Number JP23K03411). 

\appendix

\begin{figure*}[t!]
	\centering
	\includegraphics[height=2.6in,width=6.4in]{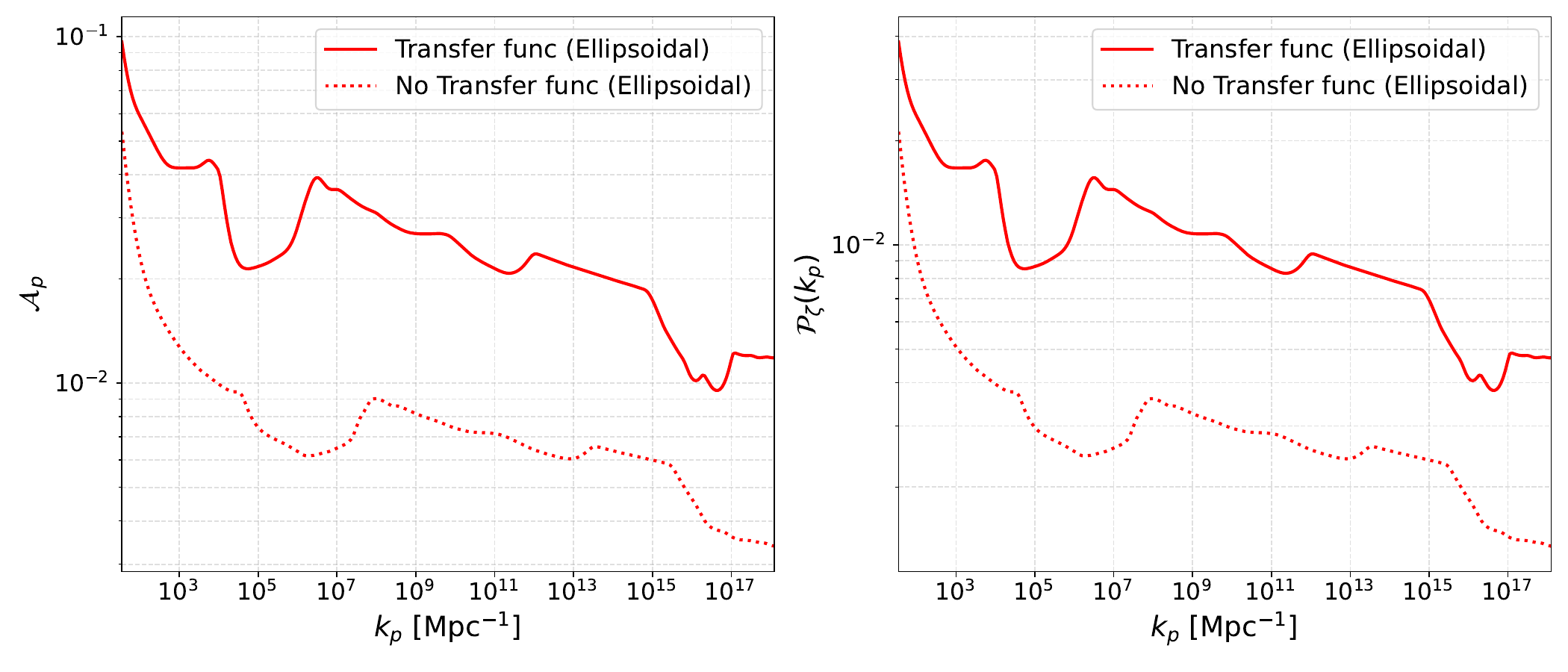} 
	\includegraphics[height=2.6in,width=6.4in]{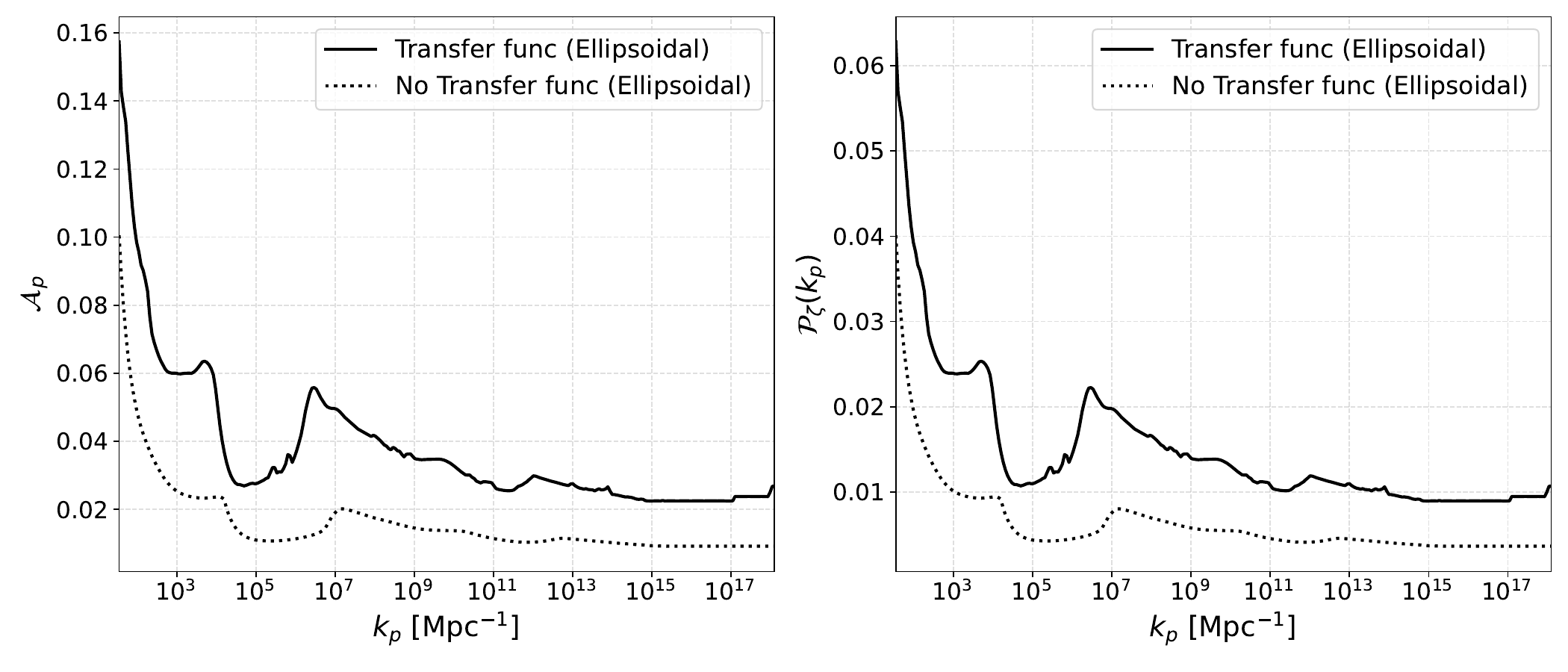} 
\caption{Effect of the use of transfer function $T (k,r_m)$ with real space top-hat window function for broad power spectrum case for peak theory (top panel) and Press-Schechter (bottom panel). The effect is shown on the constraints on $\mathcal{A}_p$ and $\mathcal{P}_{\zeta} (k_p)$ for ellipsoidal overdensity case, for broad peak case where we use $\Delta=1$ and $\alpha=0.2$. We take $\delta_c=0.45$ for the real-space top-hat window function and $\delta_c=0.18$ for the Gaussian window function. For the dotted curves, the oscillatory features at large $k_p$ arise from numerical artifacts associated with the rapidly oscillating $j_1^2(kr_m)$ term appearing in the integrals for $\sigma_L^2$ and $\mu_L^2$ in Eqns.\eqref{sigma-l} and \eqref{mu-l}. These oscillations are largely suppressed in the peak-theory case due to an effective cancellation between the $\sigma_L^2$ and $\mu_L^2$ contributions in Eq.~\eqref{beta-peak-theory-final}. In contrast, they are more pronounced in the Press-Schechter formalism, where the abundance depends directly on $\sigma_L^2$ through Eq.~\eqref{beta-ps-final}.
}
\label{appfig:mc-pt-tophat}
\end{figure*}
\begin{figure*}[t!]
\centering
\includegraphics[height=2.6in,width=6.4in]{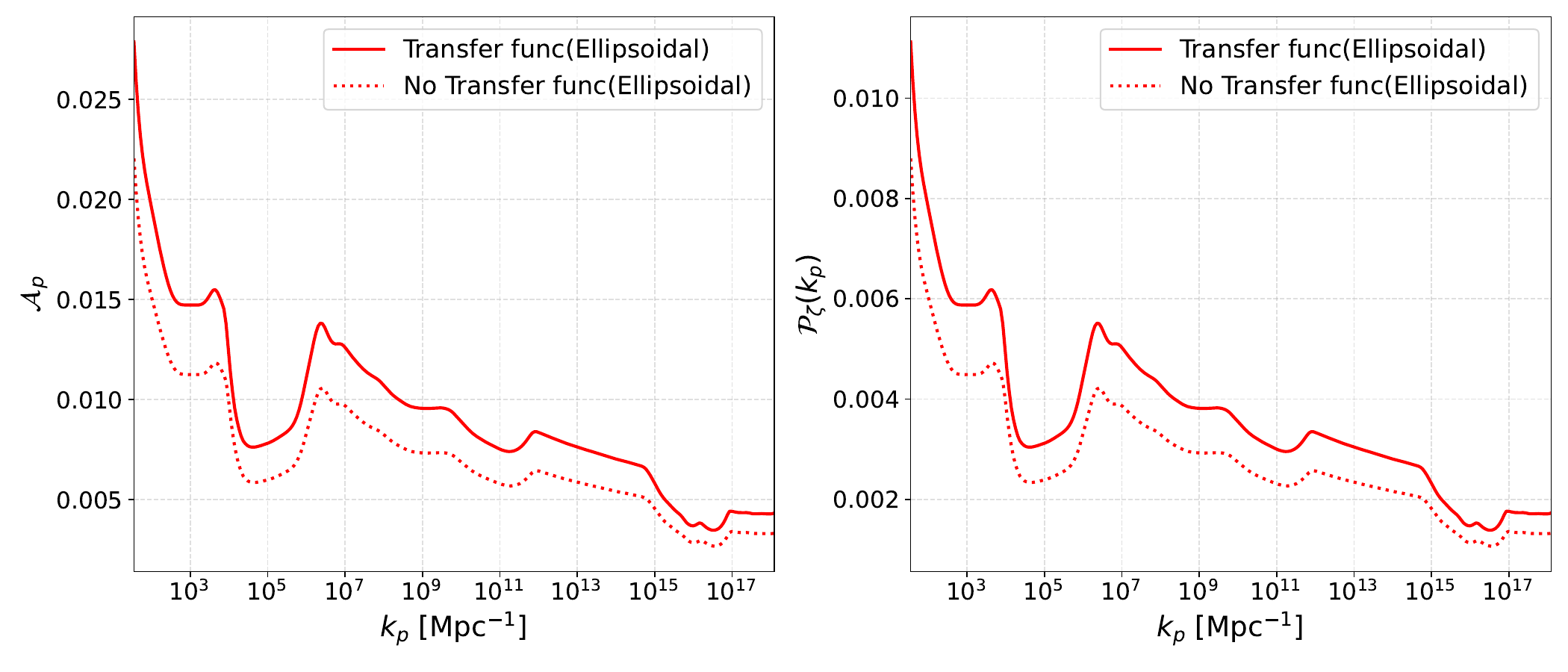} 
\includegraphics[height=2.6in,width=6.4in]{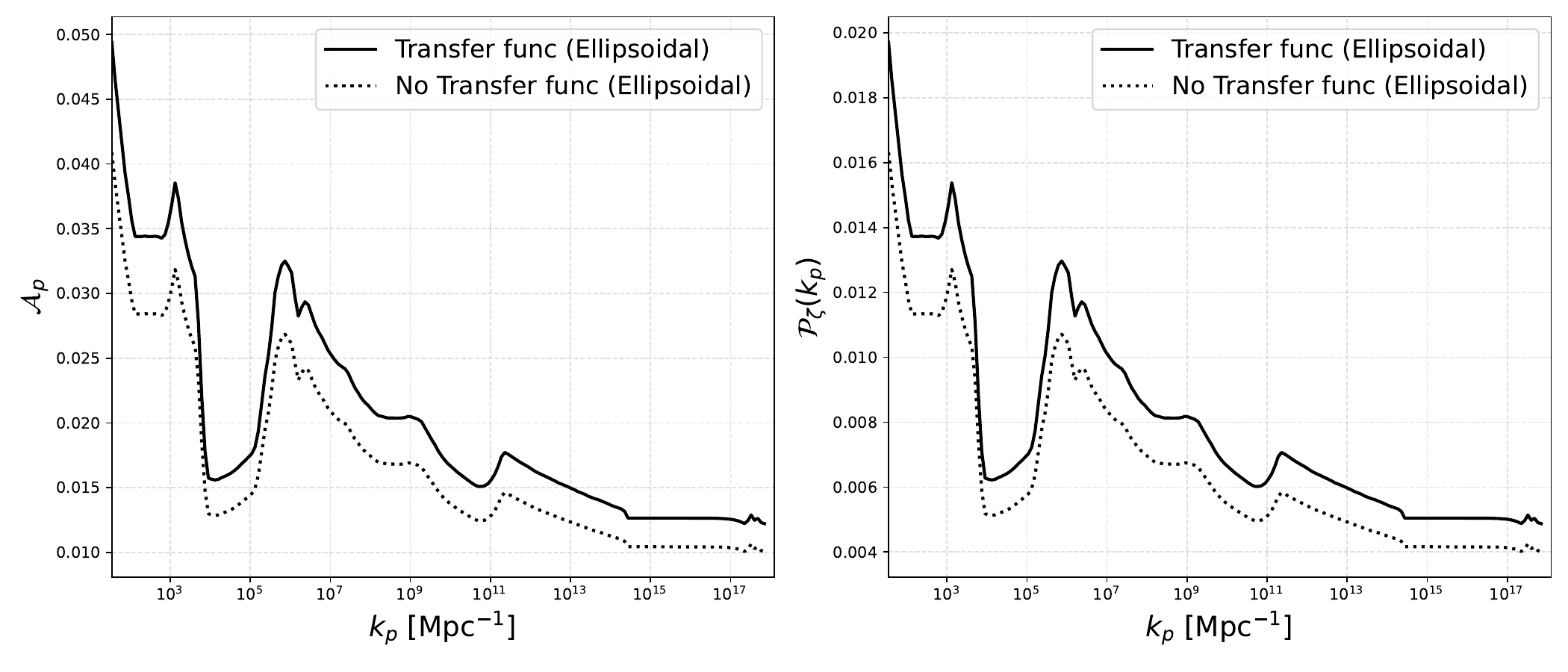} 
\caption{Effect of the use of transfer function $T (k,r_m)$ with Gaussian window function for for broad power spectrum case for peak theory (top panel) and Press-Schechter (bottom panel). The effect is shown on the constraints on $\mathcal{A}_p$ and $\mathcal{P}_{\zeta} (k_p)$ for ellipsoidal overdensity case, for broad peak case where we use $\Delta=1$ and $\alpha=0.2$.
}
\label{appfig:borad-pt-Gaussian}
\end{figure*}

\section{Non-linear relation between $\delta_m$ and $\zeta$ in the presence of non-sphericity }
\label{appendix-nr}
The relation between the averaged density contrast $\delta_m$ and $\zeta$
given by Eq.~(\ref{delta-rho-nl-flrw})
assumes spherical symmetry (i.e., $\zeta = \zeta(r)$). 
In reality, 
there is a small non-spherical correction for high-peak configurations.
When a non-spherical component is added to $\zeta$, the relation (\ref{delta-rho-nl-flrw}) needs to be reconsidered. 
To investigate the impact of non-sphericity, 
we model the non-sphericity as an ellipsoid and add a tiny non-spherical component to $\zeta$ around the peak as
\begin{equation}
\zeta = \zeta_0(r) + \epsilon \sigma_{ij} x^i x^j,
\end{equation}
where the peak is at the origin and 
a condition $\sum_i \sigma_{ii} = 0$ is imposed to make the 
second term represent purely non-spherical component. 
$\epsilon$ is a book-keeping parameter to parametrize the magnitude of the non-spherical contribution. 

With this setup, the averaged density contrast is given by
\begin{equation}
\delta_m = -\frac{3(1 + w)}{5 + 3w} \frac{1}{(aH)^2} \frac{1}{\frac{4\pi}{3} R^3} \int_{\text{ellipsoid}} e^{-2\zeta} \left( \Delta \zeta + \frac{1}{2} (\mathbf{\nabla}\zeta)^2 \right) dV.
\end{equation}
To evaluate the integral, we first decompose it as
\begin{align}
\label{appendix-I}
I &\equiv \int_{\text{ellipsoid}} e^{-2\zeta} \left( \Delta \zeta + \frac{1}{2} (\mathbf{\nabla}\zeta)^2 \right) dV \nonumber \\
&= \int_{\text{spherical}} e^{-2\zeta} \left( \Delta \zeta + \frac{1}{2} (\mathbf{\nabla}\zeta)^2 \right) dV + \left( \int_{\text{ellipsoid}} - \int_{\text{spherical}} \right) e^{-2\zeta} \left( \Delta \zeta + \frac{1}{2} (\mathbf{\nabla}\zeta)^2 \right) dV,
\end{align}
where the subscript 'spherical' denotes a sphere centered at the same origin.
Up to first order in $\epsilon$, the second term may be written as
\begin{equation}
\left( \int_{\text{ellipsoid}} - \int_{\text{spherical}} \right) e^{-2\zeta} \left( \Delta \zeta + \frac{1}{2} (\mathbf{\nabla}\zeta)^2 \right) dV = \left. e^{-2\zeta_0} \left( \Delta \zeta_0 + \frac{1}{2} (\mathbf{\nabla}\zeta_0)^2 \right) \right|_{r=b} \Delta V = 0,
\end{equation}
where $\Delta V$ is the volume difference between the sphere and the ellipsoid.
Explicitly, this can be evaluated as \cite{Kuhnel:2016exn}
\begin{equation}
\Delta V = \frac{4\pi}{3} abc - \frac{4\pi}{3} b^3 = \frac{4\pi}{3} b \left[ b \left( (c - b) - (b - a) \right) - (b - a)(c - b) \right],
\end{equation}
where $a$, $b$, and $c$ are the lengths of the principal axes of the ellipsoid.
Using $c - b = b - a$ \cite{Kuhnel:2016exn}, 
the first term vanishes. 
The second term is of second order in ellipticity, 
and hence $\Delta V = 0$ within our current approximation.
As a result, only the first term in Eq.~(\ref{appendix-I}) remains. 
To evaluate the first term, we take the principal axes of the ellipsoid to $x$, $y$, and $z$ axes. Then,
\begin{equation}
\zeta = \zeta_0(r) + \epsilon (\sigma_x x^2 + \sigma_y y^2 + \sigma_z z^2),
\end{equation}
where $\sigma_x, \sigma_y, \sigma_z$ are eigenvalues of $\sigma_{ij}$. Writing $dV$ as $dV = r^2 e^{3\zeta} (1 + r \zeta_{,r}) dr d\Omega$ and substituting these expressions, it is straightforward to compute the volume integral of $I$ up to first order in $\epsilon$. The result is given by
\begin{equation}
I = 2\pi r_b^2 e^{\zeta_0} \zeta_0' (2 + r_b \zeta_0') + \epsilon \frac{2\pi}{3} (\sigma_x + \sigma_y + \sigma_z) e^{\zeta_0} r_b^3 \left( 4 + 6 r_b \zeta_0' + r_b^2 \zeta_0'^2 \right).
\end{equation}
From the traceless condition $\sum_i \sigma_{ii} = 0$, the first-order
correction in $I$ vanishes.
Thus, up to the first order in ellipticity, there is no correction due to ellipticity to the relation between $\delta_m$ and $\zeta$ and we may use the original relation (\ref{delta-rho-nl-flrw}).

\section{Calculations on relating volume average density contrast with smoothed density contrast}
In this appendix, we show that the volume average and smoothing (with top-hat window function in real space) of the density contrast (i.e., $\delta\rho / \bar{\rho}$) are essentially the same quantities if the radius of volume is equal to the smoothing scale.
Let us consider the relation for the volume average density contrast which is defined as the volume average of $\delta\rho / \bar{\rho}$ over the radius of sphere of volume $R_m$ as
\begin{align}\label{app-deltam-rel}
\delta_{\rm average} (r_m,t) = \frac{1}{V_m} \int_0^{R_m} dR \, 4\pi R^2 \frac{\delta \rho}{\bar{\rho}}
\end{align}
where we integrated over the volume $V_m = \frac{4\pi R_m^3}{3}$. 
Alternatively, we can also define the above relation as the smoothing over the top-hat window function in real space in the following way
\begin{align}\label{app-delta-window-fun}
\delta_{\rm smooth} (r_m,t) = \int dR \, 4\pi R^2 \frac{\delta \rho}{\bar{\rho}} W_{\rm TH} (R_m , R) = \int dR \, 4\pi R^2 \frac{\delta \rho}{\bar{\rho}} \frac{3}{4\pi R_m^3} \Theta (R_m - R)
\end{align}
where $W_{\rm TH} (R_m , R) = \frac{3}{4\pi R_m^3} \Theta (R_m - R)$ is the top-hat window function in real space, and $\Theta (R_m-R)$ is the Heaviside theta function with the following values: $\Theta (R_m-R) = 1$ for $R < R_m$ and $\Theta (R_m-R) = 0 $ for $R > R_m$. Therefore, the values of the Theta function constrain the integration range to be $0 < R < R_m$ and we obtain~\eqref{app-delta-window-fun} as
\begin{align}\label{app-delta-window-fun-1}
\delta_{\rm smooth} (r_m,t) = \int_0^{R_m} dR \, 4\pi R^2 \frac{\delta \rho}{\bar{\rho}} \frac{3}{4\pi R_m^3} = \frac{1}{V_m} \int_0^{R_m} dR \, 4\pi R^2 \frac{\delta \rho}{\bar{\rho}} = \delta_{\rm average} (r_m,t)
\end{align}
where we substituted $\Theta (R_m-R) = 1$, and used Eq.\eqref{app-deltam-rel} in the last equality. Therefore, in our case both are the same and are denoted as $\delta_m = \delta_{\rm average} (r_m)= \delta_{\rm smooth} (r_m)$.
\begin{figure*}[t!]
\centering
\includegraphics[height=2.6in,width=6.4in]{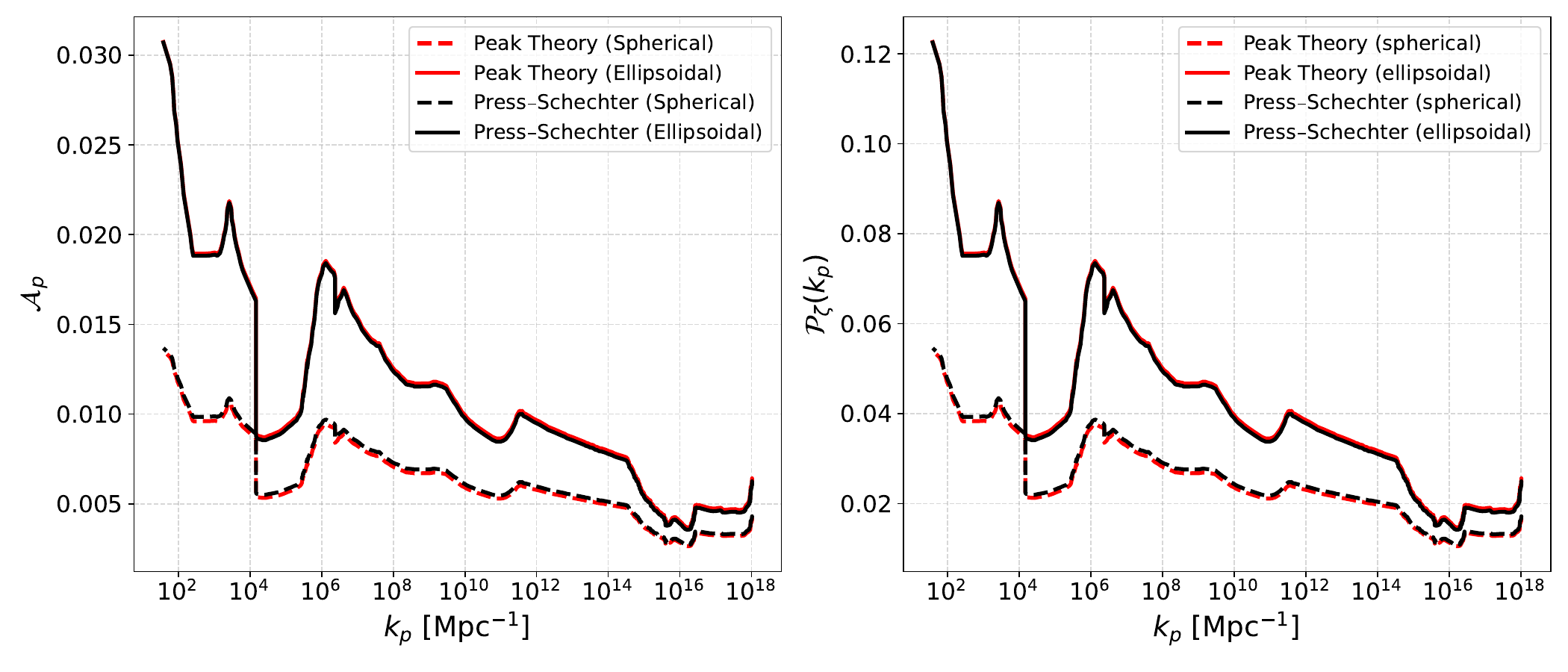} 
\caption{Comparison between Press-Schechter and peak theory on the constraints on $\mathcal{A}_p$ and $\mathcal{P}_{\zeta} (k_p)$ for both spherical and ellipsoidal overdensity case, we use $\Delta=0.1$, $\alpha=0.2$ and $\delta_c = 0.18$. The constraints are obtained from the PBHs abundance against DM, $f_{PBH}$.}
\label{appfig:Ap-Pzeta-PS-pt-top-hat-narrow}
\end{figure*}
\begin{figure*}[t!]
\centering
\includegraphics[height=2.6in,width=6.4in]{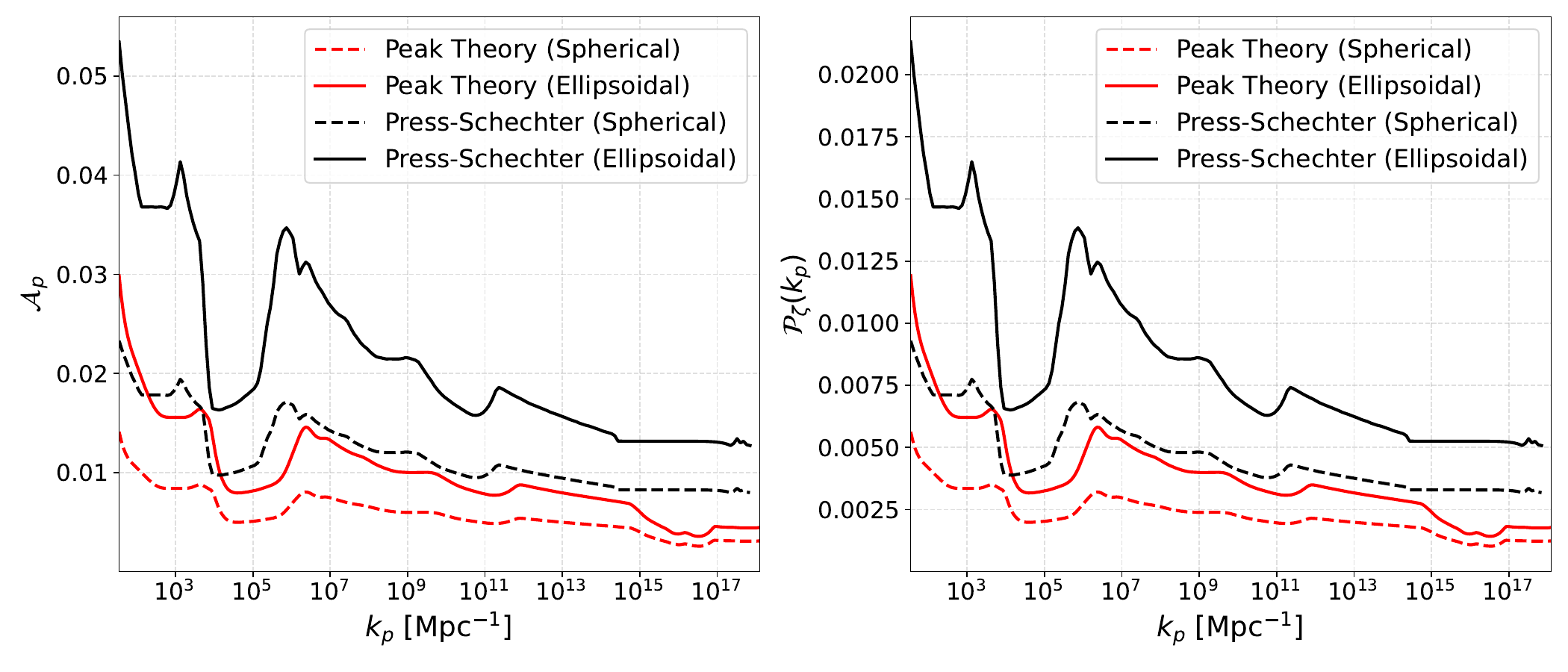} 
\caption{Comparison between Press-Schechter and peak theory on the constraints on $\mathcal{A}_p$ and $\mathcal{P}_{\zeta} (k_p)$ for both spherical and ellipsoidal overdensity case, for broad peak case where we use $\Delta=1$, $\alpha=0.2$ $\delta_c = 0.18$. The constraints are obtained from the PBHs abundance against DM, $f_{PBH}$. The apparent flattening of the Press-Schechter constraints for $10^{15}<k<10^{17}$ may be a numerical artifact arising from the highly oscillatory integration kernel in the evaluation of $\sigma_L^2$. The physical behaviour is expected to be similar to that found in the peak-theory case.}
\label{appfig:Ap-Pzeta-PS-pt-gausswindow-broad}
\end{figure*}
\begin{figure*}[t!]
\centering
\includegraphics[height=4in,width=6in]{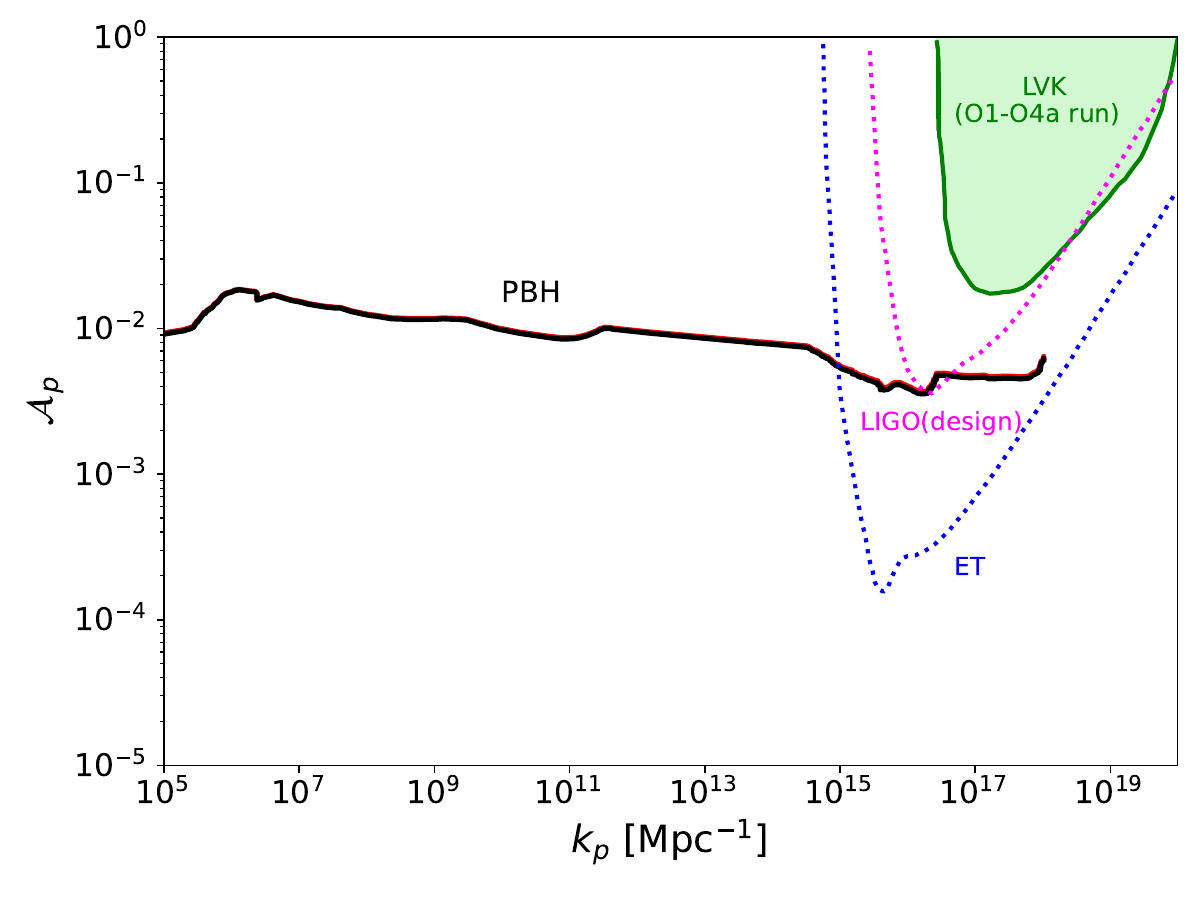} 
\caption{Comparison of our constraints for a narrow power spectrum (ellipsoidal case) from Fig.~\ref{appfig:Ap-Pzeta-PS-pt-top-hat-narrow} with other observational limits. The red and black curves show the constraints for peak theory and PS formalism, respectively. The solid Green line shows the region constrained by LVK $O1-O4a$ run~\cite{LIGOScientific:2025kry}. The dotted magenta and blue curves show the
projected sensitivities of the LIGO in its final phase (LIGO design) and the Einstein Telescope (ET), curves are taken from~\cite{Romero-Rodriguez:2021aws}. 
}
\label{ap-fig:Ap-constraint-comp-narrow-top-hat}
\end{figure*}

\section{Review of methodology for the constraints on extended PBHs mass function}
\label{sec-extended-mass}
In this appendix, we review the method proposed by Carr et al.~\cite{Carr:2017jsz} for constraining the broad-peak power spectrum, which leads to the production of PBHs with a relatively broad mass range, referred to as the extended PBH mass function. We consider an astrophysical observable $A[\psi (M)]$ depending on the PBH abundance, which can be expanded as
\begin{align}\label{A-psi-def}
A[\psi(M] &= A_0 + \int dM \, \psi (M) K_1 (M) 
+ \int dM_1 dM_2 \,  \psi (M_1)  \psi (M_2) K_2 (M_1, M_2) + ...,
\end{align}
where $\psi (M)$ is the PBH mass function, 
$A_0$ is the background contribution and the functions $K_i$ depdend on the details of the physics of observations. Assuming that the PBHs of different mass contribute independently to the astrophysical observable, 
we only need to consider the first two terms in Eq.\eqref{A-psi-def}. Thus, for the observational upper bound on the observable $A_{\rm obs}$ for a monochromatic mass function with $M_{\rm PBH} = M_c$, the PBHs mass function becomes
\begin{align}\label{psi-monoc}
\psi_{\rm mono} (M_{\rm PBH}) = f_{\rm PBH}(M_c) \, \delta (M_{\rm PBH}-M_c)~,
\end{align}
where $f_{\rm PBH}$ denotes the total PBH abundance relative to dark matter. This leads to the relation
\begin{align}\label{fpbh-mc-1}
f_{\rm PBH} (M_c) \leq \frac{A_{\rm obs}-A_0}{K_1 (M_c)} \equiv f_{\rm max} (M_c)~.
\end{align}
In the above equation $f_{\rm max}$ is the observationally allowed maxima of the fraction of DM in PBHs for a monochromatic mass distribution. Using Eq.\eqref{fpbh-mc-1} in Eq.\eqref{A-psi-def}, we have
\begin{align}\label{gen-rel-extended-mass}
\int dM \frac{\psi (M)}{f_{\rm max} (M)} \leq 1 ~,
\end{align}
which allows us to obtain the constraints on the arbitrary mass function $\psi (M)$ 
for a given $f_{\rm max}$.
However, considering each of these $N$ observables independently, the above procedure allows us to combine all the constraints by using the following relation
\begin{align}\label{gen-rel-extended-mass-all}
\sum_{i=1}^{N} \left( \int dM \frac{\psi (M)}{f_{\rm max, i} (M)} \right)^2 \leq 1 ~,
\end{align}
where $f_{\rm max, i}$ corresponds to different bounds for monochromatic mass function as defined by Eq.\eqref{fpbh-mc-1}. 
Thus, the above relation provides the constraints on the arbitrary PBH mass function.
However, we can instead use the combined monochromatic constraints from different observations provided in Ref.\cite{2021-Carr.etal-Rept.Prog.Phys} over the entire mass range and use a more simplified relation given by
\begin{align}\label{extended-mass-all-main}
\left( \int dM \frac{\psi (M)}{f_{\rm max} (M)} \right)^2 \leq 1 ~.
\end{align}
We use the above relation to derive the constraints on the broad power spectrum.
In the above equation, the numerator and denominator in integrand are provided by the theoretical model and observational bounds, respectively.

\section{Uncertainties in the estimation of constraints due to the choice of window function and transfer function}
\label{appsec-window-fun}
In this section, we discuss the effect of the inclusion of the transfer function with window function on the estimation of the constraints on the primordial power spectrum. We consider two typical cases of window functions, first is the Gaussian window function whose Fourier transform is given by
\begin{align}
\tilde{W}_G(k,r_m) = \exp\left( -\frac{(k\,r_m)^2}{4} \right)~.
\end{align}
Note that an alternative convention frequently adopted in the literature uses a Gaussian window function whose exponent differs by a factor of two from the definition adopted here~\cite{Ando:2018qdb}.In this work, however, we follow the convention used in Ref.~\cite{Gow:2020bzo,Young:2019osy}.
Second is the real-space top-hat Window function which in Fourier space is given by
\begin{align}
\tilde{W}_{\rm TH}(k,r_m) = 3\frac{\sin{(kr_m)} - kr_m \cos{(kr_m)}}{(kr_m)^3} = 3\frac{j_1 (kr_m)}{kr_m},
\end{align}
and $j_1$ is the spherical Bessel function. 
The transfer function is given by
\begin{align}
T (k,r_m) = 3\frac{\sin{\left(\frac{kr_m}{\sqrt{3}} \right)} - \left(\frac{kr_m}{\sqrt{3}} \right) \cos{\left(\frac{kr_m}{\sqrt{3}} \right)}}{\left(\frac{kr_m}{\sqrt{3}} \right)^3} = 3\frac{j_1 \left(\frac{kr_m}{\sqrt{3}} \right)}{\left(\frac{kr_m}{\sqrt{3}} \right)}~~.
\end{align}
It has been emphasized in the literature~\cite{Gow:2020bzo,Ando:2018qdb} that the real-space top-hat window function can lead to significant difficulties in the estimation of PBH abundances. In Fourier space, $\tilde{W}_{\rm TH}(k,r_m)$ suppresses high-$k$ modes less efficiently than the Gaussian window function, making the calculation more sensitive to sub-horizon fluctuations. To account for the evolution of these modes after horizon re-entry and to suppress their contribution, the transfer function $T(k,r_m)$ is commonly included in the calculation.
\begin{figure*}[t!]
\centering
\includegraphics[height=4in,width=6in]{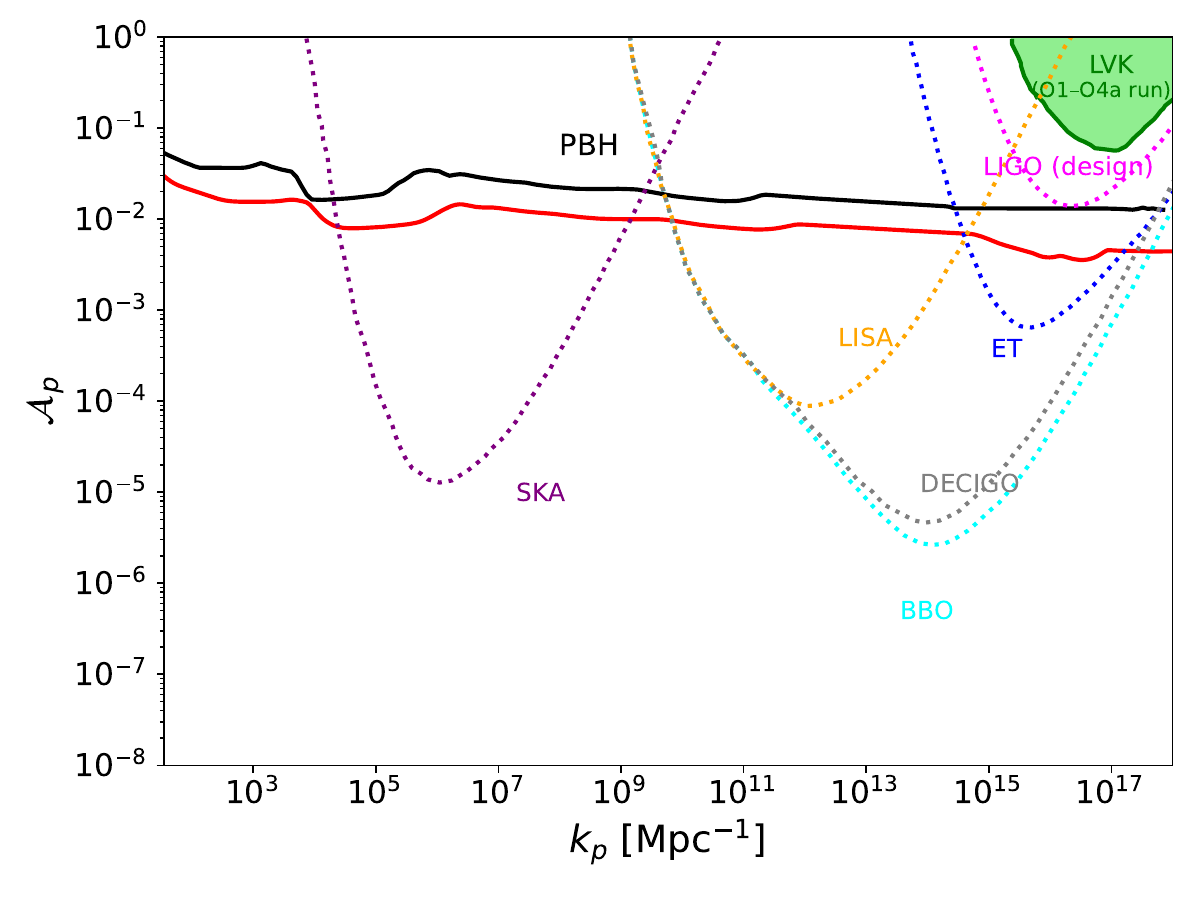} 
\caption{Comparison of our constraints for a broad power spectrum (ellipsoidal case) from Fig.~\ref{appfig:Ap-Pzeta-PS-pt-gausswindow-broad} with other observational limits. The red and black curves show the constraints for peak theory and PS formalism, respectively.
The solid Green line shows the region constrained by LVK $O1-O4a$ run~\cite{LIGOScientific:2025kry}. The dotted curves show the
projected sensitivities of the LIGO in its final phase (LIGO design) and the ET, SKA, BBO, DECIGO, LISA, which are taken from~\cite{Romero-Rodriguez:2021aws,Inomata:2018epa}. }
\label{ap-fig:Ap-constraint-comp-broad-gausswindow}
\end{figure*}
Since this effect is expected to be most relevant for extended power spectra, we focus here on the broad peak case with $\Delta=1$. Figure~\ref{appfig:mc-pt-tophat} shows the resulting constraints obtained using the real-space top-hat window function, both with and without the transfer function, for the peak-theory and Press--Schechter formalisms. We find that neglecting the transfer function leads to substantial deviations in the inferred constraints, particularly at small scales (large $k_p$). The discrepancy becomes increasingly pronounced toward larger $k_p$, demonstrating that the contribution of sub-horizon modes can significantly affect the estimated PBH abundance. Once the transfer function is included, these discrepancies are largely removed, leading to stable and physically consistent constraints. This highlights the crucial role of the transfer function when the real-space top-hat window function is employed, especially for broad primordial power spectra.
For comparison, Fig.~\ref{appfig:borad-pt-Gaussian} shows the corresponding results obtained using the Gaussian window function. In this case, the inclusion of the transfer function does not lead to dramatic changes in the shape of the constraints. The Gaussian window already provides an efficient suppression of high-$k$ modes, reducing the sensitivity to sub-horizon fluctuations. Consequently, the main effect of the transfer function is to produce a modest shift in the inferred amplitude constraints, yielding slightly weaker upper limits on the primordial power spectrum. This behavior is observed in both the Press-Schechter and peak-theory approaches.
These results demonstrate that the impact of the transfer function depends strongly on the choice of window function. While its inclusion is essential for obtaining reliable constraints with the real-space top-hat window function, it mainly produces quantitative corrections for the Gaussian window function.
\section{Constraints on the primordial power spectrum for Gaussian window function}
\label{appsec:constraints-tophat}
In this appendix, we provide the constraints on the primordial power spectrum for the case with Gaussian window function (with transfer function). To make the analysis consistent, we use the same mathedology as in the main text (for the Gaussian window function), and provide the constraints with same set of parameters. The constraints for the narrow power spectrum are shown in Figures~\ref{appfig:Ap-Pzeta-PS-pt-top-hat-narrow} and \ref{ap-fig:Ap-constraint-comp-narrow-top-hat} and for the broad power spectrum are shown in Figures~\ref{appfig:Ap-Pzeta-PS-pt-gausswindow-broad} and \ref{ap-fig:Ap-constraint-comp-broad-gausswindow}.

\bibliography{References}
\end{document}